\documentstyle[aps,twocolumn,floats,prl]{revtex}
\input epsf
\renewcommand{\l}{{\bf l}}
\newcommand{\la}{{{\bf l}_1}}
\newcommand{\lb}{{{\bf l}_2}}

\newcommand{\be}{\hat{\bf e}}
\newcommand{\len}{\phi}
\newcommand{\bn}{\hat{\bf n}}
\newcommand{\bx}{{\bf x}}
\newcommand{\cmb}{\Theta}
\newcommand{\intl}[1]{\int {d^2 {\bf l}_#1 \over (2\pi)^2}}
\newcommand{\intln}{\int {d^2 {\bf l}\over (2\pi)^2}}
\newcommand{\Ylm}[1]{Y_{l_#1}^{m_#1}}
\newcommand{\spin}[1]{\,{}_{#1}^{\vphantom{m}}}

\newcommand{\Yslm}[3]{\spin{#1} Y_{#2}^{#3}}
\newcommand{\mplus}{{\bf m_{+}}}
\newcommand{\mminus}{{\bf m_{-}}}
\newcommand{\Mp}{(\mplus \otimes \mplus)}
\newcommand{\Mn}{(\mminus \otimes \mminus)}

\newcommand{\TT}{{\Theta\Theta}}
\newcommand{\EE}{{E E}}
\newcommand{\BB}{{B B}}
\newcommand{\TE}{{\Theta E}}
\newcommand{\XP}{{X \phi}}
\newcommand{\TP}{{\Theta \phi}}
\newcommand{\EP}{{E \phi}}
\newcommand{\PP}{{\phi\phi}}
\newcommand{\Rflat}{R}
\newcommand{\Rall}{R}
\newcommand{\perm}{{\rm perm.}}
\newcommand{\btheta}{\mbox{\boldmath $\alpha$}}

\newcommand{\deld}{\delta}
\newcommand{\wj}{\left(
                          \begin{array}{ccc}
                          l_1  &  l_2  & l_3 \\
                            0  &  0    &  0
                          \end{array}
                          \right)}
\newcommand{\wjmp}[3]{\left(
                           \begin{array}{ccc}
         l_#1 & l_#2  & l_#3  \\
         m_#1 & m_#2  & m_#3
                           \end{array}
                   \right)}
\newcommand{\wjma}[6]{\left(
                           \begin{array}{ccc}
         #1 & #2  & #3  \\
         #4 & #5  & #6
                           \end{array}
                   \right)}

\begin{document}

\twocolumn[\hsize\textwidth\columnwidth\hsize\csname
@twocolumnfalse\endcsname

\title{Weak Lensing of the CMB: A Harmonic Approach}

\author{Wayne Hu}
\address{Institute for Advanced Study, Princeton, NJ 08540 \\
Revised \today}
\maketitle
\begin{abstract}Weak lensing of CMB anisotropies
and polarization for the power spectra and
higher order statistics can be handled directly in 
harmonic-space without recourse to real-space 
correlation functions.  For the power spectra,
this approach not only simplifies the calculations
but is also readily generalized from the 
usual flat-sky approximation to the exact all-sky
form by replacing Fourier harmonics with spherical harmonics.
Counterintuitively, due to the nonlinear nature
of the effect, errors in the flat-sky approximation 
do not improve on smaller scales.  They remain at the
10\% level through the acoustic regime and are sufficiently
large to merit adoption of the all-sky formalism.  
For the bispectra, a cosmic variance limited
detection of the correlation with secondary anisotropies 
has an order of magnitude greater signal-to-noise 
for combinations involving magnetic parity polarization than 
those involving the temperature alone.
Detection of these bispectra
will however be severely noise and foreground limited even with
the Planck satellite, leaving room for improvement with
higher sensitivity experiments.   We also provide a general
study of the correspondence between flat and all sky potentials,
deflection angles, convergence and shear for the power spectra and 
bispectra.
\end{abstract}
\vskip 0.5truecm
]

\section{Introduction}

As the cosmic microwave background (CMB) photons propagate 
from the last scattering surface
through intervening large-scale structure, they are gravitationally lensed.
Weak lensing effects on the the temperature and polarization distributions
of the cosmic microwave background is already a well-studied field. 
As in other aspects of the field, early work treating the effects on the 
temperature correlation function
\cite{PreRef} has largely been superceded by harmonic space
power spectrum analyses in the post-COBE era \cite{Sel96,ZalSel98}. 
In harmonic space, the physical processes of anisotropy 
formation are most directly manifest.  
However for weak lensing in the CMB, correlation function underpinnings
have typically remained, forcing transformations 
between real and Fourier space to define the effect in a 
small-angle (flat-sky) approximation.  
Exceptions include recent work on the non-Gaussianity of the lensed 
temperature field where a direct harmonic space approach has 
been taken \cite{GolSpe99,Zal99}.

In this paper, we provide a complete framework for the study of lensing
effects in the temperature and polarization fields directly in harmonic 
space.  Not only does this greatly simplify the power spectrum calculations 
but it also establishes a clear link between weak lensing power spectrum
observables in wide-field galaxy surveys and CMB observables for cross-correlation
studies.  Furthermore, this approach is easily generalized to lensing on
the full sky by replacing Fourier harmonics with spherical harmonics.

We show that counterintuitively, corrections from employing an exact all-sky
treatment are not confined to large angles. The second order 
nature of the effect brings in large scale power through mode
coupling.  
Since the all-sky expressions are as simple
to evaluate as their flat-sky approximations, which themselves are much simpler
to evaluate than the correlation function analogues, they should be employed
where full accuracy is required, e.g. for the analysis
of precise measurements from CMB satellite missions.  

Beyond the power spectrum, lensing induces three point 
correlations in the CMB through its correlation with secondary
anisotropies \cite{GolSpe99,Zal99},
 even when the intrinsic distribution at 
last scattering is Gaussian.   Detection of these effects in
the temperature maps however are severely limited by cosmic variance.
The primary anisotropies themselves act at as Gaussian
noise for these purposes.   In this case, the low level at which the CMB
is polarized can be an asset not a liability.  
Three point correlations involving the polarization, where
orientation plays a role,  are most
simply considered with their harmonic space analogue, the
bispectrum. We introduce polarization and polarization-temperature 
bispectra and show that they can have signal-to-noise advantages 
over those involving the temperature alone.  

The outline of the paper is as follows.  In \S \ref{sec:preliminaries},
we treat the basic elements of the cosmological framework, CMB temperature
and polarization, and weak lensing needed to understand these effects.
Detailed derivations are presented 
in a series of Appendices: 
\ref{sec:alllensing} covers the all-sky weak lensing
approach,
\ref{sec:wigner} the evaluation of the all-sky formulae and
\ref{sec:correspondence} the correspondence between the flat and all sky
approaches for scalar, vector and tensor fields on the sky.
The lensing effects on the power spectrum are treated in the flat-sky
approximation in \S \ref{sec:flat} and in the exact all-sky approach in
\S \ref{sec:all}.  In \S \ref{sec:bispectra}, we study the effects of
lensing on the bispectra of the temperature and polarization distributions.
We conclude in \S \ref{sec:discussion}.

\section{Formalism}
\label{sec:preliminaries}
In this section, we review and develop the formalism necessary
for calculating lensing effects in the CMB.   We review the
relevant properties of the adiabatic cold dark matter (CDM) model
in \S \ref{sec:model}. In \S \ref{sec:cmb},
we discuss the power spectra and bispectra of the temperature
fluctuations, polarization and temperature-polarization 
cross correlation. Finally in \S \ref{sec:lensing}, we 
review the properties of weak lensing relevant for the
CMB calculation.  

\subsection{Cosmological Model}
\label{sec:model}     

We work in the context of the adiabatic CDM 
family of models, where structure forms through the gravitational
instability of the CDM in a background
Friedmann-Robertson-Walker metric.  In units of the critical density 
$3H_0^2/8\pi G$, where $H_0=100h$ km s$^{-1}$ Mpc$^{-1}$ is the
Hubble parameter today, 
the contribution of each component is denoted $\Omega_i$,
$i=c$ for the CDM, $b$ for the baryons, $\Lambda$ for the cosmological
constant.
It is convenient to define
the auxiliary quantities $\Omega_m=\Omega_c+\Omega_b$ and  $\Omega_K=1-\sum_i \Omega_i$, which represent
the matter density and 
the contribution of spatial curvature to the expansion rate respectively. The expansion 
rate 
\begin{equation}
H^2 = H_0^2 \left[ \Omega_m(1+z)^3 + \Omega_K (1+z)^2 
		+\Omega_\Lambda \right]\,.
\end{equation} 
then determines the comoving conformal distance to redshift $z$,
\begin{equation}
D(z) = \int_0^z {H_0 \over H(z')} dz' \,,
\end{equation}
in units of the Hubble distance today $H_0^{-1} =2997.9h^{-1} $Mpc.
The comoving angular diameter distance 
\begin{equation}
D_A = \Omega_K^{-1/2} \sinh (\Omega_K^{1/2} D)\,,
\end{equation}
plays an important role in lensing.  Note that as $\Omega_K \rightarrow 0$, $D_A \rightarrow D$.

The adiabatic CDM model possesses a 
power spectrum
of fluctuations in the gravitational potential $\Phi$
\begin{equation}
\Delta_{\Phi}^{2}(k,z) 
= {k^3 \over 2\pi^2} P_{\Phi} = A(z)\left({k \over H_0}\right)^{n-1}T^2(k) \,,
\end{equation}
where the the transfer function is normalized to  $T(0)=1$.  
We employ the CMBFast code \cite{SelZal96}
to determine $T(k)$ at intermediate scales and extend
it to small scales using analytic fits \cite{EisHu99}.

The cosmological Poisson equation
relates the power spectra of the potential and density perturbations $\delta$
\begin{equation}
\Delta_\Phi^2 = {9 \over 4} \left( {H_0 \over k} \right)^4
\left(1 + 3{H_0^2 \over k^2}\Omega_K \right)^{-2} \Omega_m^2 (1+z)^{2}\Delta_\delta^2\,, 
\label{eqn:poisson}
\end{equation}
and gives the relationship between their relative normalization
\begin{equation}
A(z) = {9 \over 4} 
\left(1+ 3 {H_0^2 \over k^2} \Omega_K \right)^{-2} 
\Omega_m^2 F(z) \delta_H^2  \,.
\label{eqn:normalization}
\end{equation}
Here $\delta_H$ is the amplitude of present-day density fluctuations at the Hubble scale;
we adopt the COBE normalization for $\delta_H$ \cite{BunWhi97}.
$F(z)/(1+z)$ is the growth rate of linear density perturbations
$\delta(z)=F(z)\delta(0)/(1+z)$ \cite{Pee80}
\begin{equation}
F(z) \propto (1+z)
	{H(z) \over H_0} \int_z^\infty dz' (1+z') \left( {H_0 \over H(z')} \right)^3\,.
\end{equation}
For the matter dominated regime where $H \propto (1+z)^{3/2}$, 
$F$ is independent of redshift.

Although we maintain generality in all derivations, we 
illustrate our results with a $\Lambda$CDM model.  
The parameters for this model
are $\Omega_c=0.30$, $\Omega_b=0.05$, $\Omega_\Lambda=0.65$, $h=0.65$, 
$Y_p = 0.24$, $n=1$, and $\delta_H=4.2 \times 10^{-5}$.
This model has mass fluctuations on the $8 h$ Mpc$^{-1}$
scale in accord with the abundance of galaxy clusters
$\sigma_8=0.86$.  A reasonable value here
is important since the lensing calculation is second order.

\begin{figure}[t]
\centerline{\epsfxsize=3.5truein\epsffile{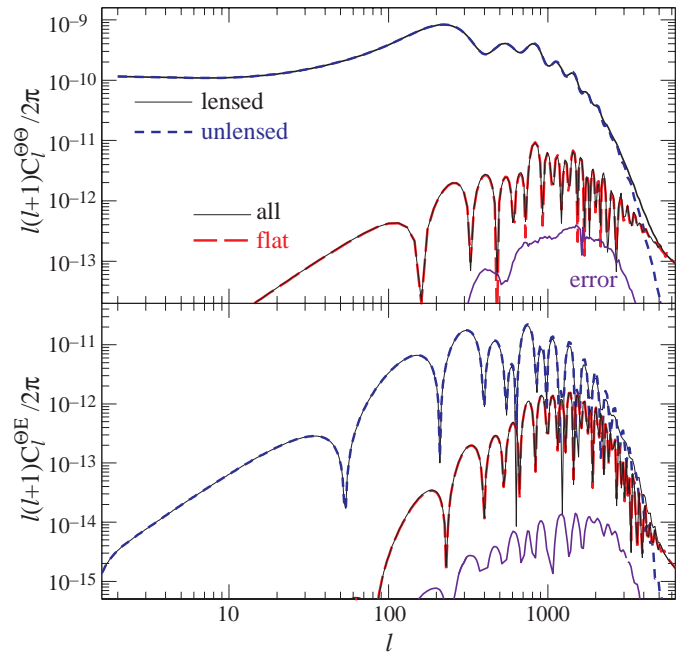}}
\caption{Temperature and temperature-polarization cross power spectra.
Shown here are the power spectra of the unlensed and lensed fields, their
difference in the all-sky and flat-sky calculations and the error induced
by using the flat sky expressions.   The oscillatory nature of the 
difference indicates that lensing smooths the power spectrum.} 
\label{fig:temp}
\end{figure}

\begin{figure}[t]
\centerline{\epsfxsize=3.5truein\epsffile{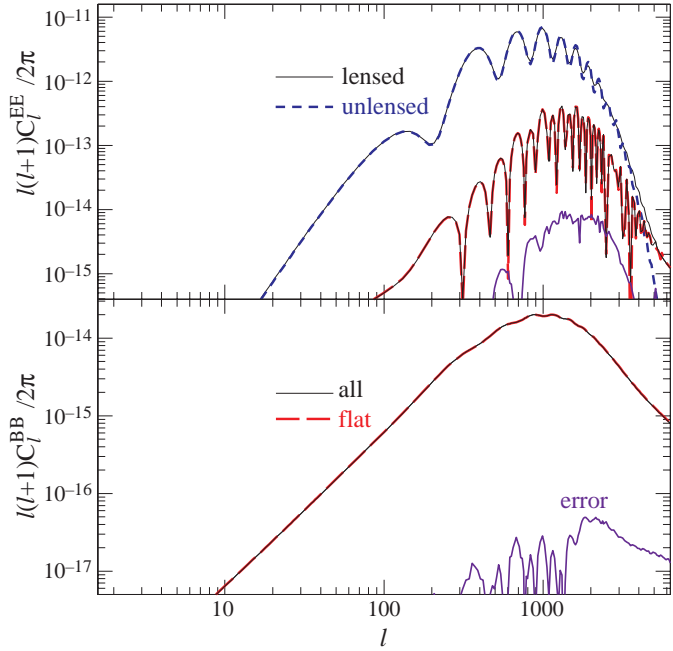}}
\caption{Polarization power spectra. The same as in Fig.~\protect{\ref{fig:temp}}
except for the $E$ and $B$ polarization.  We have assumed that the unlensed $B$
spectrum vanishes as appropriate for scalar perturbations.}
\label{fig:pol}
\end{figure}

\subsection{CMB}
\label{sec:cmb}

We decompose the CMB temperature perturbation on the sky $\Theta(\bn)=
\Delta T(\bn)/T$ into its multipole moments
\begin{equation}
   \cmb(\bn) = \sum_{l m} \cmb_{l m} Y_l^m (\bn) \,.
\end{equation}
The polarization on the sky is represented by the trace-free
symmetric Stokes matrix on the sky
\begin{eqnarray}
   {\bf P}(\bn)&=& {}_+ X(\bn)\, \Mp  
+ {}_- X(\bn)\, \Mn \,,
\end{eqnarray}
where 
\begin{eqnarray}
{}_{\pm} X(\bn) & = & Q(\bn) \pm i U(\bn) \,, \nonumber\\
    {\bf m}_\pm & = & {1 \over \sqrt{2}}(\be_\theta \mp i \be_\phi) \,.
\end{eqnarray}
The complex Stokes parameter ${}_\pm X$ is a spin-2
object which can be decomposed in the spin-spherical harmonics
\cite{Goletal67}
\begin{equation}
	{}_{\pm} X(\bn) = \sum_{l m} {}_\pm X_{lm} \spin{\pm 2}Y_{l}^{m}(\bn)\,.
\end{equation}
We have assumed that the Stokes $V$ parameter vanishes as appropriate
for cosmological perturbations; for a full set add the
term $V \epsilon_{i j}$ to the polarization matrix,
where $\epsilon_{i j}$ is the Levi-Civita tensor.

Due to the parity properties of the spin-spherical harmonics 
\begin{equation}
\spin{s} Y_l^m \rightarrow (-1)^l \spin{-s} Y_l^m \,,
\end{equation}
one introduces the parity eigenstates \cite{KamKosSte97,ZalSel97}
\begin{equation}
{}_{\pm} X_{lm} = E_{lm} \pm i B_{lm} \,,
\end{equation}
such that $E_{lm}$ just like $\Theta_{lm}$ has parity $(-1)^l$ 
(``electric'' parity) whereas $B_{lm}$ has parity $(-1)^{l+1}$
(``magnetic'' parity).  Density (scalar) fluctuations in
linear theory only stimulate the $E$ component of polarization.

The power spectra and cross correlation of these quantities 
is defined as 
\begin{equation}
    \left< X_{l m}^{*} X_{l' m'}' \right>
    = \delta_{l,l'}\delta_{m,m'} C_{l}^{XX'} \,,
\label{eqn:Clangular}
\end{equation}
where $X$ and $X'$ can take on the values $\Theta$,$E$,$B$.
Note that the cross power spectra between 
$B$ and $\Theta$ or $E$  have odd total parity and thus
vanish assuming anisotropy formation is a parity
invariant process.

The bispectrum is defined as 
\begin{equation}
    \left< X_{l m} X_{l' m'}' X_{l'' m''}'' \right>
    =   \wjma{l}{l'}{l''}{m}{m'}{m''}B_{l l' l''}^{XX'X''} \,,
\end{equation}
and vanishes if the fluctuations are Gaussian.
Even in the presence of non-Gaussianity due to non-linear
but parity-conserving sources, bispectra involving an even 
number of magnetic parity terms (including zero)
vanish for $l+l'+l''=$odd
and those involving an odd number vanish for $l+l'+l''=$even.

For a small section of the sky or high multipole moments, 
it is sufficient to treat
the sky as flat.  
In the flat-sky approximation, the Fourier moments of
the temperature fluctuations are given as
\begin{eqnarray}
\cmb(\bn) & = & \intln \cmb(\l) e^{i \l \cdot \bn} \,,
\label{eqn:fourierdecomp}
\end{eqnarray}
and the polarization as
\begin{equation}
{}_\pm X(\bn) =- \intln {}_\pm X(\l) e^{\pm 2i(\varphi_l-\varphi)}
		e^{i \l \cdot \bn} \,,
\end{equation}
where $\varphi_l$ is azimuthal angle of $\l$.  Again one separates
the Stokes moments as
\begin{equation}
{}_\pm X(\l) = E(\l) \pm i B(\l) \,.
\end{equation}

As in the all-sky case, the power spectra and cross correlations can
be defined as
with power spectra
\begin{eqnarray}
\left< X^*(\l) X'(\l') \right> &=& (2\pi)^2 \deld(\l - \l') C^{XX'}_{(l)}   \,,\\ 
\left< X^*(\l) X'(\l') X''(\l'') \right> &=& (2\pi)^2 \deld(\l - \l'-\l'') 
	B^{XX'X''}_{(\l,\l',\l'')}   \,. 
\nonumber
\end{eqnarray}
The power spectra for the fiducial $\Lambda$CDM model are shown in
Figs.~\ref{fig:temp} and \ref{fig:pol}.

In Appendix \ref{sec:correspondence}, we establish the correspondence
between the all-sky and flat-sky
spectra. 
For the power spectra and bispectra,
\begin{eqnarray}
C_l^{XX'} &\approx& C_{(l)}^{XX'}\,, \nonumber\\
B_{l l' l''}^{X X' X''} &\approx&  
			\wjma{l}{l'}{l''}{0}{0}{0}
			\sqrt{ (2l+1)(2l'+1)(2l''+1) \over 4\pi} \nonumber\\
	   &&\quad B^{X X' X''}_{(\l,\l',\l'')}\,,
\label{eqn:correspondence}
\end{eqnarray}
for sufficiently high $l$'s.

For the bispectra, we have assumed that the triplet is 
composed of an even number of magnetic parity ($B$) objects 
such that it vanishes for $l+l'+l''=$ odd. For 
combinations involving an odd number (e.g. $B\Theta\Theta$), 
the Wigner-3$j$ symbol should be replaced with its algebraic
approximation (\ref{eqn:w3j000}) but with $l+l'+l''=$ even terms set to zero
instead. However the overall sign depends on the orientation of 
the triangle in the flat-sky approximation since the bispectrum 
is then antisymmetric to reflections about either axis.

\subsection{Weak Lensing}
\label{sec:lensing}

In the so-called Born approximation where lensing effects are evaluated 
on the the null-geodesics of the unlensed photons, all effects 
can be conveniently encapsulated in the projected potential
\cite{Kai98,BarSch99}
\begin{equation}
    \len(\bn) = -2 \int dD g_{\phi}(D) \Phi(\bx(\bn),D)\,,
\label{eqn:projectedpot}
\end{equation}
where 
\begin{equation}
g_{\phi}(D) = {1 \over D_A(D)} \int_D^\infty dD' \, 
		{D_A(D'-D) \over D_A(D')} g_{s}(D')\,.
\label{eqn:lensingeff}
\end{equation}
For the CMB, the source distribution $g_{s}$ 
is the Thomson visibility
and may be replaced by a delta function at the last scattering
surface  $D_{s} = D(z \sim 10^3)$; for galaxy weak lensing this is the
distance distribution of the sources.
We explicitly relate this quantity to the more familiar convergence
and shear in Appendix \ref{sec:alllensing}.  Note that
the deflection angle is given by the angular gradient \btheta$(\bn)=
\nabla \len (\bn)$.

\begin{figure}[t]
\centerline{\epsfxsize=3.5truein\epsffile{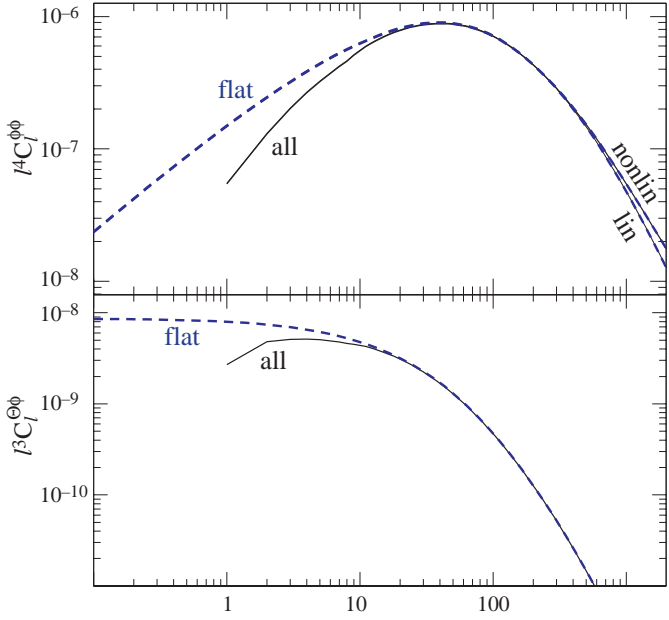}}
\caption{Lensing power spectra.  The power spectrum of the lensing potential
is shown in the top panel as calculated by the flat and all sky approaches for 
linear and non-linear density perturbations.  In the lower panel, the cross correlation
with the ISW effect is shown.  In both cases, a non-negligible fraction of the
power comes from scales where the flat-sky approximation is inadequate.}
\label{fig:lensingpower}
\end{figure}

As with the temperature perturbations, we can decompose the
lensing potential into multipole moments 
\begin{equation}
	\len(\bn) = \sum_{l m} \len_{l m} Y_{l}^{m} (\bn)\,,
\end{equation}
or Fourier moments as 
\begin{equation}
\len(\bn)  =  \intln \len(\l) e^{i \l \cdot \bn}\,,
\label{eqn:fourierdecomplen}
\end{equation}
The power spectra of the lensing potential in 
the all-sky and flat-sky cases as 
\begin{eqnarray}
    \left< \phi_{l m}^{*} \phi_{l' m'} \right>
    &=& \delta_{l,l'}\delta_{m,m'} C_{l}^{\PP} \nonumber\,,\\
\left< \phi^*(\l) \phi(\l') \right> &=& (2\pi)^2 \deld(\l - \l') C^{\PP}_{(l)}
	\,,  
\label{eqn:lenpower}
\end{eqnarray}
where again 
$C^{\PP}_{(l)} = C_{l}^\PP$.  
The lensing potential also develops a bispectrum in the non-linear density
regime,
\begin{eqnarray}
    \left< \phi_{l m} \phi_{l' m'} \phi_{l'' m''}\right>
    &=&   \wjma{l}{l'}{l''}{m}{m'}{m''}B_{l l' l''}^{\phi\phi\phi} 
\,, \nonumber\\
\left< \phi(\l) \phi(\l') \phi(\l'') \right> &=& (2\pi)^2 \deld(\l - \l'-\l'') 
	B^{XX'X''}_{(\l,\l',\l'')}    \,,
\end{eqnarray} 
which is responsible for skewness in convergence maps and other higher order
effects.  Since the lensing potential is not affected by non-linearity
until very high multipoles (see Fig.~\ref{fig:lensingpower}), we neglect
these terms here.

Finally, the lensing potential can also be correlated with 
secondary temperature and polarization anisotropies \cite{GolSpe99,SelZal99}, 
so that one must also consider the cross power spectra
\begin{eqnarray}
    \left< X_{l m}^{*} \phi_{l' m'} \right>
    &=& \delta_{l,l'}\delta_{m,m'} C_{l}^{\XP} \nonumber\,,\\
\left< X^*(\l) \phi(\l') \right> &=& (2\pi)^2 \deld(\l - \l') C_{(l)}^{\XP}
\,,
\label{eqn:crosspower}
\end{eqnarray}
in the all and flat sky limits.

To calculate the power spectra of the lensing potential for a given cosmology
one expands the gravitational potential in eqn.~(\ref{eqn:projectedpot})
in plane waves and then expanding the plane waves in spherical harmonics.
The result is
\begin{eqnarray}
        C_l^{\PP} 
         &=& 4\pi \int {dk \over k} \Delta_\Phi^2(k,z) [I_l^{\rm len}(k)]^2 \,,
\end{eqnarray}
where
\begin{eqnarray}
I_\ell^{\rm len}(k)& =&
                \int d D\, W^{\rm len}(D)
                 j_l({k \over H_0} D)  \,,\nonumber\\
W^{\rm len}(D)& =&
                -2 F(D) {D_A (D_s-D) \over
                D_A(D) D_A(D_s)}\,.
\label{eqn:lensint}
\end{eqnarray}
For curved universes, replace the spherical Bessel function
with the ultra-spherical Bessel function.
In the small scale limit, this expression may be replaced by its
equivalent Limber approximated integral \cite{Kai98}
\begin{eqnarray}
 	C_{(l)}^{\PP} &=&
           { 2 \pi^2 \over l^3 } \int d D D_A
        [W^{\rm len}(D)]^2
        \Delta^2_\Phi(k,0) 
	\Big|_{k=l {H_0 \over D_A}} 
		\,, \nonumber
\end{eqnarray}
This expression also has the useful property that its non-linear analogue
can be calculated with the replacement
\begin{eqnarray}
F(D)^2 \Delta^2_\Phi(k,0) \rightarrow  \Delta^2_\Phi(k,D) \,,
\end{eqnarray}
where the time-dependent non-linear
power spectrum is given by the scaling formula \cite{PeaDod96} and the Poisson equation (\ref{eqn:poisson}).
Since non-linear effects generally only appear at small angles,
the full non-linear all-sky spectrum can be obtained by matching
these expressions in the linear regime (see Fig.~\ref{fig:lensingpower}). 

Similarly, the cross correlation may be calculated for any secondary
effect once its relation to the gravitational potential is known.
We shall illustrate these results with the integrated Sachs-Wolfe effect.
It contributes to temperature fluctuations as 
\begin{equation}
\Theta^{\rm ISW}(\bn) = -2 \int dD\, \dot \Phi(\bx(\bn),D)\,.
\end{equation}
It then follows that the all-sky cross correlation is given by
\cite{GolSpe99,SelZal99}
\begin{eqnarray}
        C_l^{\TP} 
         &=& 4\pi \int {dk \over k} \Delta_\Phi^2(k) I_l^{\rm len}(k) I_l^{\rm ISW}(k) \,, 
\end{eqnarray}
where
\begin{eqnarray}
I_l^{\rm ISW}(k)& =&
                \int d D\, W^{\rm ISW}(D)
                 j_l({k \over H_0} D)  \,,\nonumber\\
W^{\rm ISW}(D)& =&
                -2 \dot F(D)\,,
\label{eqn:iswint}
\end{eqnarray}
again with the understanding that one replaces the spherical Bessel function
with the ultra-spherical Bessel functions for curved universes.
Similarly the flat-sky expression becomes,
\begin{eqnarray}
 	C_{(l)}^{\TP} &=&
           { 2 \pi^2 \over l^3 } \int d D D_A
        W^{\rm ISW}(D) W^{\rm len}(D)
        \Delta^2_\Phi (k) 
	\Big|_{k=l {H_0 \over D_A}} 
	\,. \nonumber
\end{eqnarray}
Figure~\ref{fig:lensingpower} also shows the 
cross-correlation for the $\Lambda$CDM
cosmology.

Cross lensing-CMB bispectrum terms can also included 
but require an external measure of lensing (e.g. a galaxy weak lensing
survey) to be observable with three-point correlations.

\section{Flat-Sky Power Spectra}
\label{sec:flat}

In this section, we calculate the effects of lensing
on the CMB temperature (\S \ref{sec:flattemp}), polarization and cross
(\S \ref{sec:flatpol}) power spectra.  The simplicity of the resulting
expressions have calculational and pedagogical advantages over
the traditional flat-sky correlation function approach \cite{Sel96,ZalSel98}.
However we also show why one cannot expect a flat-sky approach to 
be fully accurate even on small scales. 

\subsection{Temperature}
\label{sec:flattemp}

Weak lensing of the CMB remaps the primary anisotropy according
to the deflection angle $\nabla\len$
\begin{eqnarray}
\tilde\cmb(\bn) & = &  \cmb(\bn + \nabla\len) \nonumber\\
	& = &
\cmb(\bn) + \nabla_i \len(\bn) \nabla^i \cmb(\bn) \nonumber\\
&& \quad + {1 \over 2} \nabla_i \len(\bn) \nabla_j \len(\bn) \nabla^{i}\nabla^{j} \cmb(\bn)
+ \ldots 
\end{eqnarray}
Because surface brightness is conserved in lensing 
only changes the distribution of the anisotropies and has
no effect on the isotropic part of the background.

The Fourier coefficients of the lensed field then become
\begin{eqnarray}
\tilde \cmb(\l) 
&=& \int d \bn\, \tilde \cmb(\bn) e^{-i \l \cdot \bn} \nonumber\\
&=& \cmb(\l) - \intl{1} \cmb(\la) L(\l,\la)\,,
\label{eqn:thetal}
\end{eqnarray}
where
\begin{eqnarray}
L(\l,\la) &=& \len(\l-\la) \, (\l - \la) \cdot \la
+{1 \over 2} \intl{2} \len(\lb) \\ &&\quad 
\label{eqn:lfactor}
\times \len^*(\lb + \la - \l) \, (\lb \cdot \la) 
		(\lb + \la - \l)\cdot
			     \la \,.  \nonumber
\end{eqnarray}
This determines the lensed power spectrum
\begin{eqnarray}
\label{eqn:power}
\left< \tilde \cmb^*(\l) \tilde \cmb (\l') \right> & = &
	(2\pi)^2 \deld(\l - \l') \tilde C_l^\TT  \,,
\end{eqnarray}
as
\begin{eqnarray}
\tilde C_l^{\TT} &=& \left( 1
			- l^2 \Rflat \right)
			C_l^\TT  
	+ \intl{1} C_{|\l - \l_1|}^\TT C^\PP_{l_1} 
\label{eqn:ttflat}
\\ && \quad  \times
		[(\l - \la)\cdot \la]^2  \nonumber\,,
\end{eqnarray}
where
\begin{equation}
\Rflat= {1 \over 4\pi} \int {d l \over l} \, l^4 C^\PP_{l} \,.
\label{eqn:Rflat}
\end{equation} 
The second term  in eqn.~(\ref{eqn:ttflat})
represents a convolution of the power spectra.
Since $l^4 C_l^\PP$ peaks at low $l$'s compared with 
the peaks in the CMB (see Fig.~\ref{fig:lensingpower}), it can be considered as a
narrow window function on $C_l^{\TT}$ in the acoustic regime
$200 \lesssim l \lesssim 2000$. 
It is useful to consider the limit that $C_l^\TT$ is slowly
varying. It may then 
be evaluated at $\l -\la \approx \l$ and taken out of the integral
\begin{eqnarray}
	C_l^\TT \intl{1} C^\PP_{l} (\l \cdot \la)^2  &\approx&
        l^2 \Rflat C_l^\TT \,.
\end{eqnarray}
Note that the two terms in eqn.~(\ref{eqn:power}) cancel in
this limit
\begin{equation}
\tilde C_l^\TT \approx C_l^\TT \,.
\end{equation}
This is the well known result that lensing shifts but does not
create power on large scales.  Intrinsic features with width
$\Delta l$ less than the $l$ of the peak in $l^4 C_l^\PP$
are washed out by the convolution (see Fig.~\ref{fig:lensingpower}).
Note that in the $\Lambda$CDM model this scale is $l \sim 40$.
The implication is that for such a model, the 
smoothing effect even for high multipoles arises
from such low multipoles that the flat-sky approach is suspect.

On scales small compared with the damping
length $l \gtrsim 2000$, there is little intrinsic power in the CMB so that
the first term in eqn.~(\ref{eqn:ttflat}) can be ignored 
and the second term behaves instead like a smoothing of $C_l^{\phi\phi}$
of width $\Delta l$
approximately the $l$ of the peak in $l^4 C_l^\TT$. 
Since $C_l^{\phi\phi}$ is very smooth itself, the term is
approximately,   
\begin{equation}
\tilde C_l^\TT \approx C_l^\TT + {1 \over 2} l^2 C^\PP_l 
	\intl{1} l_1^2 C_{l_1}^\TT   \,,
\end{equation}
where we have interchanged the roles of $\la$ and $\la - \l$.
The power generated is proportional to the lensing
power at the same scale and may be approximated as the lensing of
a pure temperature gradient \cite{Zal99}.  In this limit the flat-sky 
approximation should be fully adequate.

\subsection{Polarization}
\label{sec:flatpol}

The lensing of the polarization field may be obtained by 
following the same steps as for the temperature field
\begin{eqnarray}
{}_{\pm} \tilde X(\bn) & = &  {}_{\pm}X(\bn + \nabla\len) \\
\label{eqn:ebl}
	& \approx &
{}_{\pm} X(\bn) + \nabla_i \len(\bn) \nabla^i {}_{\pm} X(\bn)\nonumber\\
&& \quad
+ {1 \over 2} \nabla_i \len(\bn) \nabla_j \len(\bn) \nabla^{i}\nabla^{j}
{}_{\pm}  X(\bn) \,,
\nonumber
\end{eqnarray}
where we have used the shorthand notation $\spin{\pm} X = Q\pm i U$.
The Fourier coefficients of the lensed field are then 
\begin{eqnarray}
{}_\pm \tilde X(\l) &=& {}_\pm X(\l) 
- 
\intl{1} 
{}_\pm X(\la) 
 e^{\pm 2i (\varphi_{l_1}- \varphi_{l})} L(\l,\l_1) \,,
\end{eqnarray}
where $L$ was defined in eqn.~(\ref{eqn:lfactor}).

Recalling that ${}_\pm X(\l) = E(\l)\pm i B(\l)$, we obtain the
power spectra directly 
\begin{eqnarray}
\tilde C_l^\EE &=& \left( 1- l^2 \Rflat \right) C_l^\EE + 
	 {1 \over 2} \intl{1} 
		[(\l - \la)\cdot \la]^2   
		C^\PP_{|\l - \la|} 
	\nonumber\\&&\quad \times
		[
		(C^\EE_{l_1} + C^\BB_{l_1})+ 
		\cos(4 \varphi_{l_1}) 
		(C^\EE_{l_1} - C^\BB_{l_1})] 
		\,,\nonumber\\
\tilde C_l^\BB &=& \left( 1- l^2 \Rflat \right) C_l^\BB  + 
	 {1 \over 2} \intl{1} 
		[(\l - \la)\cdot \la]^2   
		C^\PP_{|\l - \la|} 
	\nonumber\\&&\quad\times
		[
		(C^\EE_{l_1} + C^\BB_{l_1})- 
		\cos(4 \varphi_{l_1}) 
		(C^\EE_{l_1} - C^\BB_{l_1})] 
		\,,\nonumber\\
\tilde C_l^\TE &=& \left( 1- l^2 \Rflat \right) C_l^\TE + 
	  \intl{1} 
		[(\l - \la)\cdot \la]^2  
		C^\PP_{|\l - \la|} 
	\nonumber\\&&\quad\times
		C^\TE_{l_1} \cos(2 \varphi_{l_1}) 
		\,,
\end{eqnarray}
where recall that $\Rflat$ was defined in eqn.~(\ref{eqn:Rflat}).
The cross correlations between 
$B$ and $\Theta$ or $E$ still vanish since lensing is parity
conserving.  
Unlike the case of the temperature
fluctuations, lensing does not conserve the broadband large scale power of the
$E$ and $B$ \cite{ZalSel98}, but only the total polarization power.
For example, lensing will create a $B$ component in a field that originally had only
an $E$-component.  Furthermore, lensing actually
destroys temperature-polarization cross correlations due to
the lack of correlation with the generated $B$ polarization.
From Fig.~\ref{fig:temp}, one can see that the largest
relative effect of lensing is on the correlation.

\section{All-Sky Power Spectra}
\label{sec:all}

In this section, we treat lensing effects on the temperature
(\ref{sec:alltemp}), polarization and cross (\ref{sec:allpol})
power spectra in a full all-sky formalism.  
Corrections to the flat-sky 
results remain at the 10\% even on small scales.
Moreover, although the derivation
appears more complicated, the end results for the 
power spectra are simple.  They are as readily evaluated their
the flat-sky counterparts and should be used in their stead.

\subsection{Temperature}
\label{sec:alltemp}

In the all-sky case, the Fourier harmonics are replaced with spherical harmonics, and
the lensed field becomes
\begin{eqnarray}
\tilde \cmb_{l m} &\approx& \cmb_{l m} + \int d\bn Y_l^{m*} 
	\nabla_i \len(\bn) \nabla^i \cmb(\bn) \nonumber\\
&& \quad + {1 \over 2} 
\int d\bn Y_l^{m*} 
\nabla_i \len(\bn) \nabla_j \len(\bn) \nabla^{i}\nabla^{j} \cmb(\bn)
\nonumber\\
& = & \cmb_{l m} + \sum_{l_1 m_1} \sum_{l_2 m_2} 
			\len_{l_1 m_1}  \cmb_{l_2 m_2} 
\label{eqn:thetalm}
\\ && \quad \times			
\bigg[ I_{l l_1 l_2}^{m m_1 m_2} 
+ {1 \over 2}  
\sum_{l_3 m_3} 
			\len_{l_3 m_3}^* 
J_{l l_1 l_2 l_3}^{m m_1 m_2 m_3} \bigg] \,, \nonumber
\end{eqnarray}
with the geometrical factors expressed as integrals over the spherical harmonics
\begin{eqnarray}
I_{l l_1 l_2}^{m m_1 m_2} &=&
			\int d\bn \, Y_l^{m*} \left( \nabla_i \Ylm{1} \right)
					      \left( \nabla^i \Ylm{2} \right)\,, \\
\label{eqn:IJform}
J_{l l_1 l_2 l_3}^{m m_1 m_2 m_3}
			 &=& \int d\bn\, Y_l^{m*}  
				\left( \nabla_i \Ylm{1}        \right)
				\left( \nabla_j Y_{l_3}^{m_3*} \right)
				 \nabla^{i}\nabla^j \Ylm{2}      \,.\nonumber
\end{eqnarray}
The
lensed power spectrum then becomes
\begin{eqnarray}
\tilde C_l &=& C_l + \sum_{l_1 l_2} 
	C_{l_1}^\PP C_{l_2}^\TT S_1 
	+ C_l^\TT \sum_{l_1} C_{l_1}^\PP  S_2 \,,
\label{eqn:ttpowerlong}
\end{eqnarray}
with
\begin{eqnarray}
S_1 &=& \sum_{m_1 m_2} \left(I_{l l_1 l_2}^{m m_1 m_2}\right)^2\,,  \nonumber\\
S_2 &=& {1 \over 2}\sum_{m_1} J_{l l_1 l l_1}^{m m_1 m m_1} + {\rm cc} \,,
\end{eqnarray}
where ``cc'' denotes the complex conjugate and we have suppressed
the $l$-indices.

These formidable looking expressions simplify considerably.
The second term may be rewritten through integration by parts and the 
identity
$\nabla^2 Y_l^m = -l(l+1) Y_l^m$ \cite{GolSpe99},
\begin{eqnarray}
I_{l l_1 l_2}^{m m_1 m_2} &=& 
{1 \over 2}
[l_1(l_1+1) + l_2(l_2+1) - l (l+1)] 
\nonumber\\ && \times
\int d\bn\, Y_l^{m*}\, \Ylm{1}\, \Ylm{2}\,.
\end{eqnarray}
The remaining integral may be expressed in terms of the Wigner-3$j$ symbol
through the general relation
\begin{eqnarray}
&&
\int d\bn \left( \Yslm{s_1}{l_1}{m_1*}    \right) \Yslm{s_2}{l_2}{m_2}
	  \left( \Yslm{s_3}{l_2}{m_3} \right) =
\nonumber\\
&& \qquad (-1)^{m_1+s_1}
\sqrt{(2 l_1+1)(2 l_2 + 1)(2 l_3+1) \over 4\pi} \nonumber\\
&& \qquad \times \wjma {l_1}{l_2}{l_3}{s_1}{-s_2}{-s_3}
 \wjma{l_1}{l_2}{l_3}{-m_1}{m_2}{m_3}\,, 
\label{eqn:threey}
\end{eqnarray}
where note that $\spin{0} Y_l^m = Y_l^m$.
It is therefore convenient to define
\begin{eqnarray}
F_{l_1 l_2 l_3} &=&
{1 \over 2}[l_2(l_2+1) + l_3(l_3+1) - l_1 (l_1+1)] 
\\&&\times
\sqrt{(2 l_1+1)(2 l_2 + 1)(2 l_3 +1)  \over 4\pi} 
\wjma {l_1}{l_2}{l_3}{0}{0}{0}  \nonumber\,.
\end{eqnarray}
Finally the Wigner-3$j$ symbol obeys 
\begin{eqnarray}
\sum_{m_1 m_2} \wjmp{1}{2}{3} \wjmp{1}{2}{3} = 
{1 \over 2 l_3+1}\,.
\label{eqn:ortho}
\end{eqnarray}
Putting these relations together, we find that
\begin{equation}
S_1
	= {1 \over 2l+1} \left( F_{l l_1 l_2} \right)^2 \,.
\label{eqn:Isum}
\end{equation}
An algebraic expression for the relevant Wigner-3$j$ symbol is given
in the Appendix.

The second term in eqn.~(\ref{eqn:ttpowerlong})
can be simplified
by re-expressing the gradients of the spherical harmonics
with spin-1 spherical harmonics.   As shown in Appendix
\ref{sec:alllensing}, the spin-1 harmonics are the eigenmodes of
vector fields on the sky and naturally appear in expressions
for deflection angles.
Note that there is a general relation for raising and lowering
the spin of a spherical harmonic
\cite{Goletal67},
\begin{eqnarray}
 {\bf m}_- \cdot {\bf \nabla} \spin{s} Y_l^m &
= & \sqrt{(l-s)(l+s +1) \over 2} \spin{s+1} Y_l^m  \,,
\nonumber\\
 {\bf m}_+ \cdot {\bf \nabla} \spin{s} Y_l^m &
= & -\sqrt{(l+s)(l-s +1) \over 2} \spin{s-1} Y_l^m \,,
\label{eqn:raiselower}
\end{eqnarray}
so that
\begin{eqnarray}
\nabla Y_l^m = \sqrt{ l(l+1) \over 2} \left[ 
		{}_1 Y_l^m {\bf m_+} 
		-{}_{-1} Y_l^m {\bf m_-}
		\right] \,.
\label{eqn:ygrad}
\end{eqnarray}
As an aside, we note that equation (\ref{eqn:Isum}) can alternately
be derived from this relation and the integral (\ref{eqn:threey})
with $s=\pm 1$.

Further, we note that spin spherical harmonics also 
obey a sum rule \cite{HuWhi97}
\begin{eqnarray}
\sum_m {}_{s_1} Y_l^{m*}(\bn) \,
	{}_{s_2} Y_l^{m}(\bn)  &=& \sqrt{ 2l +1 \over 4\pi} {}_{s_2} Y_l^{-s_1}({\bf 0}) \,.
\label{eqn:sumrule}
\end{eqnarray}
For the spin-1 harmonics 
\begin{equation}
{}_{-1} Y_l^1({\bf 0}) = {}_1 Y_l^{-1}({\bf 0}) = - \sqrt{2l+1 \over 4\pi} \,,
\end{equation}
and the others involving $s_1,s_2=\pm 1$ vanish. 
These results imply that
\begin{eqnarray}
\sum_m \nabla_i Y_l^m \nabla_j Y_l^{m*} &=& {1 \over 2} l(l+1) {2l + 1 \over 4\pi} 
   [(\mplus)_i (\mminus)_j  
\nonumber\\&&\quad
+ (\mminus)_i (\mplus)_j] \,.
\end{eqnarray}
To evaluate the second derivative term in
equation~(\ref{eqn:IJform}), we again apply equation~(\ref{eqn:raiselower})
to show that 
\begin{eqnarray}
&&[(\mplus)_i (\mminus)_j  + (\mminus)_i (\mplus)_j] \nabla^i\nabla^j 
	\spin{s} Y_l^m  =
\nonumber\\ && \qquad 
 -[l(l+1)-s^2] \spin{s} Y_l^m \,.
\label{eqn:secondderiv}
\end{eqnarray}
Putting these expressions together we obtain,
\begin{eqnarray}
S_2=
- {1 \over 2} 
			l(l+1) \, l_1(l_1+1) { 2 l_1 +1 \over 4\pi} \,.
\label{eqn:Jsum}
\end{eqnarray}
Finally combining expressions eqns.~(\ref{eqn:ttpowerlong}), 
(\ref{eqn:Isum}) and
(\ref{eqn:Jsum}),
we have the following simple result
\begin{eqnarray}
\tilde C_l^\TT &=& \left[ 1 - l(l+1) \Rall \right] C_l^\TT + 
\sum_{l_1,l_2} 
	C_{l_1}^\PP C_{l_2}  {\left( F_{l l_1 l_2} \right)^2 \over 2l+1} \,,
\label{eqn:ttallsky}
\end{eqnarray}
where
\begin{equation}
\Rall= {1 \over 2} \sum_{l_1} {l_1}(l_1 +1) {2l_1+1 \over 4\pi} C_{l_1}^\PP \,.
\label{eqn:Rall}
\end{equation}
This expression is computationally no more involved than the flat-sky expression
eqn.~(\ref{eqn:ttflat}) and has the benefit of being exact.
Since the lensing effect even at high $l$ in the CMB originates from the low order 
multipoles of $\phi$, corrections due to the curvature of the sky are not 
confined to low $l$.  We show in Fig.~\ref{fig:temp} that the correction
causes a 10\% difference in the effect.  The change in 
$\tilde C_l^{\Theta\Theta}$ itself is even smaller (of order 1\%). 
Nonetheless it is larger than the cosmic variance of these high 
multipoles and thus
should be included in calculations for full accuracy.  Corrections
can be even larger in models with a red tilt $n<1$ in the
initial spectrum.
 
\subsection{Polarization}
\label{sec:allpol}

The derivation of the all-sky generalization for polarization is superficially more
involved but follows the same steps as in the temperature case and 
results in expressions that are no more difficult to evaluate. 
The lensed polarization multipoles are given by 
\begin{eqnarray}
{}_\pm X_{l m} 
& = & 
{}_\pm X_{l m}
+ \sum_{l_1 m_1} \sum_{l_2 m_2}
			\len_{l_1 m_1}  {}_\pm X_{l_2 m_2}
\\&&\quad\times
\label{eqn:eblm}
\left[ \spin{\pm 2} I_{l l_1 l_2}^{m m_1 m_2}
		  + {1 \over 2}  
			\sum_{l_3 m_3} 
			\len_{l_3 m_3}^* 
\spin{\pm 2} J_{l l_1 l_2 l_3}^{m m_1 m_2 m_3}\right]\,,\nonumber
\end{eqnarray}
with the 
geometrical factors expressed now as integrals over the spin-spherical harmonics
\begin{eqnarray}
\spin{\pm 2} I_{l l_1 l_2}^{m m_1 m_2} &=&
			\int d\bn \, \spin{\pm 2} Y_l^{m*} \left( \nabla_i \Ylm{1} \right)
					      \left( \nabla^i \spin{\pm 2} \Ylm{2} \right) 
\,,\nonumber\\
\spin{\pm 2} J_{l l_1 l_2 l_3}^{m m_1 m_2 m_3}
			 &=& \int d\bn\, \spin{\pm 2} Y_l^{m*}  
				\left( \nabla_i \Ylm{1}        \right)
				\left( \nabla_i Y_{l_3}^{m_3*} \right)
\nonumber\\&&\quad\times
				\left( \nabla^{i} \nabla^j \spin{\pm 2} \Ylm{2}     \right) \,.
\end{eqnarray}
Noting that
\begin{eqnarray}
\spin{\pm 2} I_{l l_1 l_2}^{m m_1 m_2} 
	 &=& (-1)^{L} \spin{\mp 2} I_{l l_1 l_2}^{m m_1 m_2}    \,,
\end{eqnarray}
where $L= l+l_1+l_2$ and recalling that ${}_\pm X_{lm}= E_{lm}\pm iB_{lm}$, 
the power spectra then become 
\begin{eqnarray}
\tilde C_l^\EE &=& C_l^\EE + {1 \over 2} 
	\sum_{l_1 l_2} C_{l_1}^\PP 
	   \Big[ \left(C_{l_2}^\EE + C_{l_2}^\BB \right) 
\\&&
\label{eqn:eepowerlong}
		+ (-1)^{L} 
	          \left(C_{l_2}^\EE - C_{l_2}^\BB \right) \Big]
		{}_{22} S_1
		\nonumber\\
	&& + 
	{1 \over 2} C_l^\EE 
		\sum_{l_1} C_{l_1}^\PP \left(
		{}_{2} S_2 + {}_{-2} S_2\right)\,,
\nonumber
\end{eqnarray}
where
\begin{eqnarray}
{}_{22} S_1     &=& \sum_{m_1 m_2} \left(
				\spin{2}  I_{l l_1 l_2}^{m m_1 m_2} 
		  \right)^2 \,, \nonumber\\
{}_{\pm 2} S_2 &=&  {1 \over 2}\sum_{m_1} 
			\spin{\pm 2} J_{l l_1 l_1 l}^{m m_1 m_1 m} + {\rm cc} \,.
\end{eqnarray}
The expression for $\tilde C_l^\BB$ follows by interchanging
$EE$ and $BB$.
The cross power spectrum is
\begin{eqnarray}
\tilde C_l^\TE &=& C_l^\TE + {1 \over 2} 
	\sum_{l_1 l_2} C_{l_1}^\PP 
	    C_{l_2}^\TE [1 + (-1)^{L}] {}_{02} S_1
%
\nonumber\\
	&& \quad + {1 \over 4 } C_l^\TE   
		\sum_{l_1} C_{l_1}^\PP \Big(
		\spin{2} S_2  + \spin{-2} S_2 
		+ 2 S_2 \Big)    \,,
\label{eqn:tepowerlong}
\end{eqnarray} 
with
\begin{equation}
{}_{02} S_1 = \sum_{m_1 m_2} \left(
			I_{l l_1 l_2}^{m m_1 m_2} 
				\spin{2}  
			I_{l l_1 l_2}^{m m_1 m_2} 
		\right)  \,.
\end{equation}
Just as in the case for the temperature field, these expressions simplify considerably.
The spin-$2$ harmonics are eigenfunctions of the angular Laplacian of a tensor
\begin{equation}
\nabla^2 \spin{\pm 2} Y_l^m  = [-l(l+1)+4] \spin{\pm 2} Y_l^m  \,,
\end{equation}
which follows from contracting indices in equation~(\ref{eqn:secondderiv}).
It then  follows  that
\begin{eqnarray}
\spin{\pm 2} I_{l l_1 l_2}^{m m_1 m_2}
&=& {1 \over 2}[l_1(l_1+1) + l_2(l_2+1) - l(l+1)]  
\nonumber\\ && \quad
\int d\bn \left( \Yslm{\pm 2}{l}{m*}    \right) \Ylm{1} 
	  \left( \Yslm{\pm 2}{l_2}{m_2} \right) \,.
\end{eqnarray}
Comparison with eqn.~(\ref{eqn:threey}) implies that 
it is convenient then to define the quantity
\begin{eqnarray}
\spin{2} F_{l_1 l_2 l_3} &=& {1 \over 2} [l_2(l_2+1) + l_3(l_3+1) - l_1(l_1+1)]
\\ &&
\sqrt{(2 l_1+1)(2 l_2 + 1)(2 l_3+1) \over 4\pi}\wjma{l_1}{l_2}{l_3}{2}{0}{-2} \,,
\nonumber
\end{eqnarray}
so that
\begin{eqnarray}
{}_{22} S_1 
&=& {1 \over 2l+1}(\spin{2} F_{l l_1 l_2})^2 \,, \nonumber\\
{}_{02} S_1 &=& {1 \over 2l+1}\left(F_{l l_1 l_2} \right) \left( \spin{2} F_{l l_1 l_2} \right)\,.
\end{eqnarray}

The third term in equation~(\ref{eqn:eepowerlong}) can be simplified by
following the same steps for the analogous temperature term except for
the replacement of $s=0$ with $s=\pm 2$ in equation (\ref{eqn:raiselower}).
The result is 
\begin{eqnarray}
\spin{\pm 2} S_2 
= - {1 \over 2} [l(l+1)-4] \, l_1(l_1+1) { 2 l_1 +1 \over 4\pi} \,.
\label{eqn:Jsum2}
\end{eqnarray}
Putting these relations together, we obtain the result for the power spectra
\begin{eqnarray}
\tilde C_l^\EE &=& \left[ 1 - (l^2+l-4)\Rall \right] C_l^\EE + 
{1 \over 2} \sum_{l_1,l_2} 
	C_{l_1}^\PP 
	{\left( \spin{2} F_{l l_1 l_2} \right)^2 \over 2l+1}
	\nonumber\\&& \quad
		\left[  C_{l_2}^\EE + C_{l_2}^\BB +
		(-1)^L (C_{l_2}^\EE - C_{l_2}^\BB) \right]  \,,
\nonumber\\
\tilde C_l^\BB &=& \left[ 1 - (l^2+l-4)\Rall \right] C_l^\BB + 
{1 \over 2} \sum_{l_1,l_2} 
	C_{l_1}^\PP 
	{\left( \spin{2} F_{l l_1 l_2} \right)^2 \over 2l+1}
	\nonumber\\&& \quad
		\left[  C_{l_2}^\EE + C_{l_2}^\BB -
		(-1)^L (C_{l_2}^\EE - C_{l_2}^\BB) \right] \,,
\nonumber\\
\tilde C_l^\TE &=& \left[ 1 - (l^2+l-2)\Rall \right] C_l^\TE + 
\sum_{l_1,l_2} C_{l_1}^\PP
	\nonumber\\&& \quad
	{\left(          F_{l l_1 l_2} 
		\spin{2} F_{l l_1 l_2} \right) \over 2l+1}
		 C_{l_2}^\TE  \,.
\label{eqn:polallsky}
\end{eqnarray}
Recall that $L= l+l_1+l_2$ and $R$ was defined in eqn.~(\ref{eqn:Rall}).
These expressions are plotted for the $\Lambda$CDM model in 
Fig.~\ref{fig:pol}.


\section{Flat and All Sky Bispectra}
\label{sec:bispectra}

In this section, we consider the lensing contributions to CMB bispectra
through the correlation with secondary anisotropies.
We begin by reviewing the calculations for the temperature bispectrum
as previously treated by \cite{GolSpe99,Zal99}.  We then introduce
the polarization and cross bispectra which in principle
have signal-to-noise advantages over the temperature bispectra.
We illustrate the formalism with a concrete calculation of the
effect due to the ISW secondary anisotropy.

\subsection{Temperature}
\label{sec:bitemp}

Contributions to the temperature bispectra from the cross
power spectrum $C_l^\TP$ discussed in \S \ref{sec:lensing} follow
immediately from the first order lensing term, i.e. 
eqn.~(\ref{eqn:thetalm}) for the all-sky bispectrum \cite{GolSpe99},
\begin{equation}
B_{l_1 l_2 l_3}^{\Theta\Theta\Theta} = 
	F_{l_1 l_2 l_3} C_{l_2}^\TP C_{l_3}^\TT + 5 \perm\,,
\end{equation}
and 
eqn.~(\ref{eqn:thetal}) for the flat sky bispectrum \cite{Zal99}
\begin{eqnarray}
B^{\Theta\Theta\Theta}_{(\l_1,\l_2,\l_3)} &=& 
			- (\l_2 \cdot \l_3) 
	 			C_{l_2}^\TP  C_{l_3}^\TT + 5\perm 
\end{eqnarray}
One can show that these relations satisfy the general expression 
for the correspondence between flat and all sky bispectra 
eqn.~(\ref{eqn:correspondence}) by noting that 
\begin{equation}
\l_2 \cdot \l_3  
		= -{1 \over 2}(l_2^2 + l_3^2 - l_1^2) \,,
\label{eqn:dotproduct}
\end{equation}
since the angles of a triangle is fully defined 
by the length of its sides.

Note that there can be strong cancellation between the
terms in the permutation in both cases. As we have seen, the
spectrum of $\phi$ is generally peaked to low multipoles 
implying a corresponding weighting of $C_l^\TP$ to low multipoles
for secondary anisotropies that correlate strongly with $\phi$.  
In this case the triangles $(l_1,l_2,l_3)$ that contribute
most strongly are highly flattened such that
two sides nearly coincide in length $l_1 \approx l_3 \gg l_2$.  
In this case,  contributions $l_1^2$ and $l_3^2$ 
in eqn.~(\ref{eqn:dotproduct}) 
are cancelled off the permutation $\l_3 \leftrightarrow \l_1$ leaving 
only a term of order $l_2^2$.  

These considerations also signal problems for the flat-sky
expressions.  It is important to know what on scales most of the
detectable signal is coming from. 
In the all-sky formalism, the signals from the $m$ modes are added together with
weights given by the Wigner-3$j$ symbol
\begin{eqnarray}
B_{l_1 l_2 l_3}^{\Theta\Theta\Theta} &=&
	\sum_{m_1 m_2 m_3} 
\wjma{l_1}{l_2}{l_3}{m_1}{m_2}{m_3} 
\left< \Theta_{l_1 m_1} \Theta_{l_2 m_2} \Theta_{l_3 m_3} \right> \,. \nonumber\\
\end{eqnarray}	
For the small effects due to the correlation of secondary anisotropies
with lensing, the covariance of the bispectrum estimators is dominated
by the Gaussian noise from the power spectrum \cite{Luo94} 
\begin{eqnarray}
{\rm Cov} = 
C_{l_1}^\TT C_{l_2}^\TT C_{l_3}^\TT 
		\delta_{l_1 l_1'}
		\delta_{l_2 l_2'}
		\delta_{l_3 l_3'} + 5 \perm \,,
\end{eqnarray} 
where the permutations are in the indices of the $l'$ triplet.
The overall signal-to-noise becomes 
\begin{eqnarray}
\left( {S \over N} \right)^2 = 
				\sum_{l_1 l_2 l_3}
				\sum_{l_1' l_2' l_3'}
				B_{l_1 l_2 l_3}^{\Theta\Theta\Theta}
				[{\rm Cov}^{-1}]
				B_{l_1' l_2' l_3'}^{\Theta\Theta\Theta} \,.
\end{eqnarray}			
The covariance is in general  diagonal in the 6$\times$6 blocks 		
of permutations of $(l_1,l_2,l_3)$ and for this simple case of the
temperature bispectrum, the blocks are proportional to the trivial 
matrix of all ones.  The result is one can take a simple sum over
all distinct triplets or equivalently divide the full sum by a
factor of 6,
\begin{equation}
\left( {S \over N} \right)^2 = 
				\sum_{l_1 l_2 l_3}
				{(B_{l_1 l_2 l_3}^{\Theta\Theta\Theta})^2
					\over 6 
				C_{l_1}^\TT 
				C_{l_2}^\TT
				C_{l_3}^\TT} \,,
\end{equation}
for a cosmic variance limited experiment.  For a realistic
experiment with noise from the detectors and residual foregrounds, 
one simply replaces
\begin{equation}
C_l^{XX'} \rightarrow C_l^{XX'} + C_l^{XX' {\rm (noise)}} \,,
\end{equation}
here and below. 
Note that one can also construct the Fisher information matrix of
the bispectrum along these lines \cite{CooHu99}.

Correspondingly, in the flat-sky one constructs the
optimal inverse-variance weighted statistic \cite{Zal99} (see also
Appendix \ref{sec:correspondence})
\begin{equation}
\left( {S \over N} \right)^2 = 
{f_{\rm sky} \over \pi}{1 \over (2\pi)^2} \int d^2 l_1 \int d^2 l_2
				{[B_{(l_1,l_2,l_3)}^{\Theta\Theta\Theta}]^2
					\over 6 
				C_{l_1}^\TT 
				C_{l_2}^\TT
				C_{l_3}^\TT} \,,
\end{equation}
where $f_{\rm sky}$ is the fraction of the sky covered.  We show
that these expressions are equivalent in the high $l$, $f_{\rm sky}=1$
limit in Appendix \ref{sec:correspondence}.  
Thus the extra factor of $f_{\rm sky}$ can be included in the all-sky 
expression to approximate the effects incomplete sky coverage due to
exclusion of regions contaminated by galactic foregrounds. 

The weighting of the modes is such that the quantity of
interest in the lensing-temperature correlation is $l^3 C_l^\TP$ where
the extra factor of $l$ over the straight bispectrum contribution comes
from the square root of the volume factor in $l$-space.
This quantity is plotted in Fig.~\ref{fig:lensingpower} for the
cross correlation with the ISW effect.   The implication is that
for this effect, full accuracy requires an all-sky approach and we 
shall hereafter use this to evaluate the signal-to-noise. 

\begin{figure}[t]
\centerline{\epsfxsize=3.5truein\epsffile{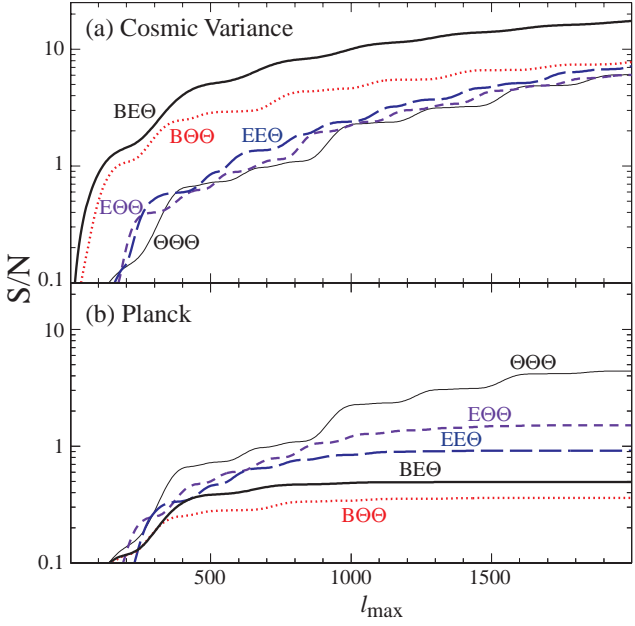}}
\caption{Cumulative signal-to-noise in the bispectra as a function of maximum $l$
for a cosmic variance limited experiment and for the Planck satellite. 
Note that for the cosmic variance limited case (a), bispectra involving
the $B$-polarization have a substantial signal-to-noise advantage over
the other bispectra.  For the Planck satellite (b), we assume that the
additional variance comes only from detector noise.  In practice,
residual foreground contamination and sky-cuts to avoid them
will lower the signal-to-noise further.
}
\label{fig:bispectrum}
\end{figure}

The overall signal-to-noise as a function of the 
largest $l$ included in the sum is shown in Fig.~\ref{fig:bispectrum}
for a cosmic variance limited experiment and the Planck satellite
(see \cite{CooHu99} for the specification of the noise).  Note
that the Planck satellite is effectively cosmic variance limited
to $l \sim 1000$ and even so the $S/N$ is only of order a few 
\cite{GolSpe99}.

\subsection{Polarization and Cross Correlation}
\label{sec:bipol}

Bispectra involving the $E$ and $B$ parity polarization will also
receive contributions from the correlation induced by lensing. 
Although these signals are smaller than the temperature bispectrum
in an absolute sense, we have seen that the main obstacle in 
detecting the temperature bispectrum is cosmic variance from
the Gaussian contributions. 

We begin by analyzing terms that do not involve the $B$-parity
polarization.  For these all-sky bispectra, only terms with
$L\equiv l_1 + l_2 + l_3$=even are non-vanishing, and we will implicitly assume
that only even terms are considered.  With the help of 
eqns.~(\ref{eqn:ebl}) and (\ref{eqn:eblm}), we can immediately
write the all and flat sky results as 
\begin{eqnarray}
B^{E\Theta\Theta}_{l_1 l_2 l_3}
	&=& 
	\spin{2} F_{l_1 l_2 l_3} 
	C_{l_2}^\TP C_{l_3}^\TE   
	\nonumber\\&&
	+ F_{l_2 l_1 l_3} 
	( C_{l_1}^\EP C_{l_3}^\TT +
	C_{l_1}^\TE C_{l_3}^\TP)
		+ (l_2 \leftrightarrow l_3) \,,
\nonumber\\
%
B^{E\Theta\Theta}_{(\l_1,\l_2,\l_3)}
	&=&
			- (\l_2 \cdot \l_3) 
			  \cos 2\varphi_{31}
			C_{l_2}^\TP C_{l_3}^\TE   
\\ &&
			- (\l_1 \cdot \l_3) 
			( C_{l_1}^\EP C_{l_3}^\TT 
+   
			C_{l_1}^\TE C_{l_3}^\TP)
			+ (\l_2 \leftrightarrow \l_3)  \,,
\nonumber
\end{eqnarray}		
where
\begin{equation}
\varphi_{A B}=\varphi_{l_A}-\varphi_{l_B} \,.
\end{equation}

The general correspondence between the flat and all sky expressions in 
eqn.~(\ref{eqn:correspondence}) is established by 
the 
use of eqn.~(\ref{eqn:dotproduct})
the approximation
discussed in the Appendix
\begin{equation}
\wjma{l_1}{l_2}{l_3}{2}{0}{-2} \approx
 	 \cos 2\varphi_{31}
	\wjma{l_1}{l_2}{l_3}{0}{0}{0}\,, 
\end{equation}
for $L=$even.
The cancellation for flattened triangles discussed in \S \ref{sec:bitemp}
still applies and is easiest to see in the flat-sky limit: the
flatness of the triangles implies $\cos 2\varphi_{31} \sim 1$.

For the $S/N$ calculation, note that the covariance is given by
\begin{eqnarray}
{\rm Cov} &=& 
		C_{l_1}^{EE} C_{l_2}^{\Theta\Theta} C_{l_3}^{\Theta\Theta}
		\delta_{l_1 l_1'} \delta_{l_2 l_2'} \delta_{l_3 l_3'} 
	\nonumber\\ && \quad +
		C_{l_1}^{\Theta E} C_{l_2}^{\Theta\Theta} C_{l_3}^{\Theta E}
		\delta_{l_1 l_2'} \delta_{l_2 l_3'} \delta_{l_3 l_1'} 
	\nonumber\\ && \quad +
		C_{l_1}^{\Theta E} C_{l_2}^{\Theta E} C_{l_3}^{\Theta \Theta}
		\delta_{l_1 l_3'} \delta_{l_2 l_1'} \delta_{l_3 l_2'}
		+ (l_2' \leftrightarrow l_3') \,,
\end{eqnarray}
so that a full calculation requires inverting this matrix for
each distinct triplet.  Since we are interested mainly in the
order of magnitude of $S/N$, we can set the lower bound as 
\begin{equation}
\left( {S \over N} \right)^2 \ge 
				\sum_{l_1 l_2 l_3}
				{(B_{l_1 l_2 l_3}^{E\Theta\Theta})^2
					\over 6 
				C_{l_1}^\EE 
				C_{l_2}^\TT
				C_{l_3}^\TT} \,,
\end{equation}
which amounts to ignoring duplicate triplets and 
replacing the remaining triplet with the average $S/N$ of the
set.
This limit is plotted for the ISW effect in Fig.~\ref{fig:bispectrum} 
as a function of the maximal $l_1$ included in the sum.  As expected, it
is comparable to the signal-to-noise in the temperature
bispectrum.  Of course, it is experimentally more difficult to
achieve the cosmic variance limit in the polarization with
a realistic experiment containing detector and foreground 
noise.

There is also a qualitatively new effect from the polarization-lensing
correlation $C_l^{E\len}$. However since secondary polarization only
arises from Thomson scattering effects, we expect this contribution 
to be small in $\Lambda$CDM models where the optical depth during
reionization is $\tau < 0.3$ \cite{CooHu99}.  

The $E\Theta\Theta$ bispectrum term is
\begin{eqnarray}
B^{E E\Theta}_{l_1 l_2 l_3} 
	&=& 
	(\spin{2} F_{l_1 l_2 l_3} 
		C_{l_2}^\EP C_{l_3}^\TE  
	+ 
	 \spin{2} F_{l_1 l_3 l_2} 
		C_{l_2}^\EE C_{l_3}^\TP
	)   
\nonumber\\  && \quad
      +
	F_{l_3 l_1 l_2} 
	C_{l_1}^\EP C_{l_2}^\TE   
			+ (l_1 \leftrightarrow l_2)   \,,
\nonumber\\
B^{E E\Theta}_{(\l_1,\l_2,\l_3)} 
	&=& 
			- (\l_2 \cdot \l_3) 
			  \Big( \cos 2\varphi_{31}
	 C_{l_2}^\EP C_{l_3}^\TE  
\nonumber\\&&\quad
			  +\cos 2\varphi_{21}
			C_{l_2}^\EE C_{l_3}^\TP\Big)   \nonumber\\
	&&\quad
			- (\l_1 \cdot \l_2) 
			C_{l_1}^\EP C_{l_2}^\TE   
			+ (\l_1 \leftrightarrow \l_2)  \,,
\end{eqnarray}	
with covariance
\begin{eqnarray}
{\rm Cov} &=& 
		C_{l_1}^{EE} C_{l_2}^{EE} C_{l_3}^{\Theta\Theta}
		\delta_{l_1 l_1'} \delta_{l_2 l_2'} \delta_{l_3 l_3'} 
	\nonumber\\ && \quad +
		C_{l_1}^{EE} C_{l_2}^{E\Theta} C_{l_3}^{\Theta E}
		\delta_{l_1 l_2'} \delta_{l_2 l_3'} \delta_{l_3 l_1'} 
	\nonumber\\ && \quad +
		C_{l_1}^{\Theta E} C_{l_2}^{E E} C_{l_3}^{\Theta E}
		\delta_{l_1 l_3'} \delta_{l_2 l_1'} \delta_{l_3 l_2'}
		+ (l_1' \leftrightarrow l_2') \,,
\end{eqnarray}
with which we can bound the $S/N$
\begin{equation}
\left( {S \over N} \right)^2 > 
				\sum_{l_1 l_2 l_3}
				{(B_{l_1 l_2 l_3}^{EE\Theta})^2
					\over 6 
				C_{l_1}^\EE 
				C_{l_2}^\EE
				C_{l_3}^\TT} \,.
\end{equation}
Again, the ISW example is shown in Fig.~\ref{fig:bispectrum}.

Finally the $EEE$ bispectrum is given by
\begin{eqnarray}
B^{E E E}_{l_1 l_2 l_3} 
	&=& 
	\spin{2} F_{l_1 l_2 l_3} 
	C_{l_2}^\EP C_{l_3}^\EE  + 5 \perm 
\nonumber\,\\
B^{E E E}_{(\l_1,\l_2,\l_3)}&=& 
			- (\l_2 \cdot \l_3) 
			  \cos 2 \varphi_{31}
			  C_{l_2}^\EP C_{l_3}^\EE  + 5 \perm
\end{eqnarray}	
with covariance
\begin{eqnarray}
{\rm Cov} = 
C_{l_1}^\EE C_{l_2}^\EE C_{l_3}^\EE
		\delta_{l_1 l_1'}
		\delta_{l_2 l_2'}
		\delta_{l_3 l_3'} + 5 \perm \,,
\end{eqnarray}
and signal-to-noise
\begin{equation}
\left( {S \over N} \right)^2 = 
				\sum_{l_1 l_2 l_3}
				{(B_{l_1 l_2 l_3}^{EEE})^2
					\over 6 
				C_{l_1}^\EE 
				C_{l_2}^\EE
				C_{l_3}^\EE} \,.
\end{equation}
This bispectrum signal vanishes for the ISW effect.

Bispectra involving the $B$-parity polarization have distinct 
properties.  For terms 
involving one B-parity polarization term, only 
$l_1+l_2+l_3$=odd contributes to the all-sky spectrum and we 
implicitly assume below that even terms vanish.

For the $B\Theta\Theta$ bispectrum,
\begin{eqnarray}
B^{B\Theta\Theta}_{l_1 l_2 l_3}
	&=& 
	i \left(\spin{2} F_{l_1 l_2 l_3} \right)
	C_{l_2}^\TP C_{l_3}^\TE   
			+ (l_2 \leftrightarrow l_3)  \,,
\nonumber\\
B^{B\Theta\Theta}_{(\l_1,\l_2,\l_3)}
	&=&
			- (\l_2 \cdot \l_3) 
			  \sin 2\varphi_{31}
			C_{l_2}^\TP C_{l_3}^\TE   
			- (\l_2 \leftrightarrow \l_3) \,.
\label{eqn:BTT}
\end{eqnarray}		

Again the correspondence between the flat and all sky expressions in 
eqn.~(\ref{eqn:correspondence}) is established by the approximation
discussed in the Appendix
\begin{equation}
\wjma{l_1}{l_2}{l_3}{2}{0}{-2} \approx
 	  \pm i \sin 2\varphi_{31}
	\wjma{l_1}{l_2}{l_3}{0}{0}{0}\,, 
\end{equation}
for $L=$odd.
The sign ambiguity comes from the fact that a reflection of the
triangle $(\l_1,\l_2,\l_3)$ across one of the axes corresponds to
remappings $\varphi \rightarrow \pi -\varphi$ or $\varphi \rightarrow -\varphi$ and
hence a reversal in sign of the flat-sky bispectrum in 
equation~(\ref{eqn:BTT}).  
In this case the cancellation for flattened triangles discussed
in \S \ref{sec:bitemp} does {\it not} apply.  However since
$\sin 2\varphi_{31} \approx 2\varphi_{31} \ll 1$, a suppression still exists.

The covariance of the $B\Theta\Theta$ bispectrum is 
\begin{eqnarray}
{\rm Cov} &=& 
		C_{l_1}^{BB} C_{l_2}^{\Theta\Theta} C_{l_3}^{\Theta\Theta}
		\delta_{l_1 l_1'} \delta_{l_2 l_2'} \delta_{l_3 l_3'} 
		+ (l_2' \leftrightarrow l_3') \,,
\end{eqnarray}
leading to a signal-to-noise
\begin{equation}
\left( {S \over N} \right)^2 = 
				\sum_{l_1 l_2 l_3}
				{(B_{l_1 l_2 l_3}^{B\Theta\Theta})^2
					\over 2 
				C_{l_1}^\BB 
				C_{l_2}^\TT
				C_{l_3}^\TT} \,.
\end{equation}
In a cosmic variance limited experiment (see Fig.~\ref{fig:bispectrum}), 
the $B\Theta\Theta$ bispectrum has signal-to-noise advantages over its 
temperature and $E$ polarization
counterparts due to the fact that for scalar perturbations
$C_{l_1}^\BB$ is dominated by the lensing contributions themselves.
Moreover, even if the tensor contributions are near their 
current limits of $T/S \lesssim  0.3$, the signal-to-noise is not
much affected for $l \gtrsim 100$ due to the strong damping of gravity
wave contributions under the horizon scale at last scattering.
However for the Planck experiment, the detection is severely limited
by detector noise and may also suffer further degradation from
incomplete foreground subtraction \cite{Tegetal99}.

Next, the $BE\Theta$ bispectrum is given by
\begin{eqnarray}
B^{B E \Theta}_{l_1 l_2 l_3}
	&=& 
	i 
	(
	  \spin{2} F_{l_1 l_2 l_3} 
		C_{l_2}^\EP C_{l_3}^\TE   
			+ 
	  \spin{2} F_{l_1 l_3 l_2} C_{l_2}^\EE C_{l_3}^\TP   
	)  \,,
\nonumber\\
B^{B E \Theta}_{(\l_1,\l_2,\l_3)}
	&=&
			- (\l_2 \cdot \l_3) 
			  \Big(
			\sin 2\varphi_{31}
				C_{l_2}^\EP C_{l_3}^\TE   
\\ && \quad 
			+
			\sin 2\varphi_{21}
	  			C_{l_2}^\EE C_{l_3}^\TP   \Big)\,,
\nonumber
\end{eqnarray}
with a covariance
\begin{eqnarray}
{\rm Cov} &=& 
		C_{l_1}^{BB} C_{l_2}^{EE} C_{l_3}^{\Theta\Theta}
		\delta_{l_1 l_1'} \delta_{l_2 l_2'} \delta_{l_3 l_3'} 
		\nonumber\\
		&&\quad +C_{l_1}^{BB} C_{l_2}^{\Theta E} C_{l_3}^{\Theta E}
		\delta_{l_1 l_1'} \delta_{l_2 l_3'} \delta_{l_3 l_2'} 
\,,
\end{eqnarray}
leading to a signal-to-noise
\begin{equation}
\left( {S \over N} \right)^2 \ge 
				\sum_{l_1 l_2 l_3}
				{(B_{l_1 l_2 l_3}^{B\Theta\Theta})^2
					\over 2 
				C_{l_1}^\BB 
				C_{l_2}^\EE
				C_{l_3}^\TT} \,.
\end{equation}
The signal-to-noise of this term can be greater than that of
$B\Theta\Theta$ due to the fact that the temperature 
and $E$-polarization are only
partially correlated in the unlensed sky.

Finally,
\begin{eqnarray}
B^{B E E}_{l_1 l_2 l_3}
	&=& 
	i 
	 \left(  \spin{2} F_{l_1 l_2 l_3} \right)
		C_{l_2}^\EP C_{l_3}^\EE   
			+ (l_2 \leftrightarrow l_3)   \,,
\\
B^{B E E}_{(\l_1,\l_2,\l_3)}
	&=&
			- (\l_2 \cdot \l_3) 
			\sin 2 \varphi_{31}
				C_{l_2}^\EP C_{l_3}^\EE   
			- (\l_2 \leftrightarrow \l_3) \,,
\nonumber
\end{eqnarray}
with a covariance
\begin{eqnarray}
{\rm Cov} &=& 
		C_{l_1}^{BB} C_2^{EE} C_3^{EE}
		\delta_{l_1 l_1'} \delta_{l_2 l_2'} \delta_{l_3 l_3'} 
		+ (l_2' \leftrightarrow l_3') \,,
\end{eqnarray}
leading to a signal-to-noise
\begin{equation}
\left( {S \over N} \right)^2 = 
				\sum_{l_1 l_2 l_3}
				{(B_{l_1 l_2 l_3}^{B E E})^2
					\over 2 
				C_{l_1}^\BB 
				C_{l_2}^\EE
				C_{l_3}^\EE} \,.
\end{equation}
This signal vanishes for the ISW effect.

Terms involving more than one $B$ term have no contributions to 
first order in the correlation power spectrum. 

\section{Discussion}
\label{sec:discussion}

We have shown that a harmonic approach to weak lensing in the CMB provides
a simple and exact means of calculating its effects on the temperature
and polarization power spectra,
given the power spectrum of the lensing
potential or convergence,  
and on the analogous bispectra given their power spectrum of
the cross correlation with secondary anisotropies.
Corrections to the flat-sky approximations appear 
even at high multipoles because
even there, lensing effects arises from the large-scale fluctuations in the deflection
angles.  These corrections correspond to a change in
the predictions at the $\mu$K level.  While this is a negligible change
given observations today, it is above the cosmic-variance limit
and should be included when interpreting the high-precision results expected
from Planck.

Unlike the temperature bispectrum, bispectra involving both the temperature 
and polarization multipoles of the CMB have the potential of producing a 
high signal-to-noise $(\sim 10)$ detection of secondary anisotropies such as 
the ISW effects even with relatively modest angular 
resolutions $l < 1000$.  Other secondary anisotropies such as the
Sunyaev-Zel'dovich effect are expected to contribute even stronger
signals, although their exact amplitude is far more uncertain
presently \cite{GolSpe99}. 

Achieving a cosmic-variance limited
detection of the magnetic-parity polarization is a daunting challenge.  
Even signal-to-noise near unity requires  detectors which 
are a factor of 3 more sensitive
to polarization than those planned for the Planck satellite.  
Also of concern are the residual foreground contamination remaining in
the maps after multifrequency subtraction.  Our current best models of
the foregrounds indicate that with the Planck channels and sensitivities,
foregrounds and detector noise may enter into the polarization maps with comparable 
amplitudes \cite{Tegetal99}.  Thus improving the actual 
sensitivity to the cosmic
signal beyond the specifications of the Planck experiment will not only
require better detectors but also a better understanding of the foregrounds,
perhaps with increased frequency coverage and sampling. 

Nonetheless, the polarization of the CMB offers the potential to open a new
window on physical processes at low redshifts and the opportunity to learn more
from the CMB than can be achieved with the next generation of CMB satellites. 
  
{\it Acknowledgements:} I would like to thank M. Zaldarriaga for many useful
discussions and help during the early stages of this work.  
This work was supported by the Keck Foundation,
a Sloan Fellowship, and NSF-9513835.

\appendix
\section{All-Sky Weak Lensing Observables}
\label{sec:alllensing}

All weak lensing observables may be defined in terms of the projected
potential $\phi$   
\begin{equation}
    \len(\bn) = -2 \int dD g_\len (D) \Phi(\bx(\bn),D)\,,
\label{eqn:projectedpota}
\end{equation}
or equivalently its multipole moments $\len_{lm}$ in the all-sky
formalism or Fourier coefficients $\len(\l)$.  Recall
from eqn.~(\ref{eqn:lensingeff})
that $g_\len$ is the lensing efficiency function.

The deflection angle that a photon suffers while traveling from the
source at $D_{s}$ is given by the angular gradient of the potential
\btheta$(\bn)=\nabla \len (\bn)$.  Applying equation~(\ref{eqn:ygrad}) to the
the spherical harmonic expansion, we obtain 
\begin{equation}
\btheta
=\sum_{l m} \sqrt{ l(l+1) \over 2 }\phi_{l m} \left[ \spin{1} Y_l^m \mplus 
			                 - \spin{-1} Y_l^m \mminus 
		\right]\,.
\end{equation}
This implies that the quantity $\alpha_1 \pm i \alpha_2$ is a 
spin $\pm 1$ object
\begin{eqnarray}
[\alpha_1 \pm i \alpha_2](\bn) 
&\equiv& \sum_{l m} ( c  \pm i g)_{l m} \spin{\pm 1} Y_l^m(\bn) \nonumber\\
&=& \pm \sum_{l m}\sqrt{l(l+1)} \len_{l m} \spin{\pm 1} Y_l^m(\bn) \,,
\end{eqnarray}
which states that the curl term $c_{lm}$ vanishes and the 
gradient term
\begin{equation}
g_{l m} = -i \sqrt{l(l+1)} \len_{l m} \,.
\end{equation}
The power spectrum of the angular deflection is then
\begin{eqnarray}
\left< g_{l' m'}^* g_{l m}\right> &\equiv&
		\delta_{l,l'} \delta_{m,m'} C_l^{gg} \nonumber\\
&=& \delta_{l,l'} \delta_{m,m'} l(l+1) C_l^{\len\len}\,,
\end{eqnarray}
with the curl power vanishing.  This 
accounts for the factors of $l(l+1)$ in equations involving
the angular deflection [e.g. eqn.~(\ref{eqn:Rall})].

The corresponding flat-sky quantity is given by the decomposition
[see eqn.~(\ref{eqn:spin1})]
\begin{eqnarray}
[\alpha_1 \pm i \alpha_2](\bn)  &\equiv& 
\pm i 
\int {d^2 l \over (2\pi)^2} 
[c \pm i g](\l) e^{\pm i(\varphi_l -\varphi)}
			e^{i \l \cdot \bn}\,,
\end{eqnarray}
with $c(\l)=0$ and
\begin{eqnarray}
g(\l) &=& - i l \len(\l)\,, \nonumber\\ 
C^{gg}_{(l)} &=& l^2 C^{\phi\phi}_{(l)} \,.
\end{eqnarray}

These relations also give the bispectrum of the deflection angle
in terms of bispectrum of the lensing potential in the obvious manner.

The convergence $(\kappa)$ and shear $(\gamma_1,\gamma_2)$ 
are familiar weak lensing observables from galaxy weak lensing studies
\cite{BarSch99}.
Although they are not directly needed for CMB studies, they are of interest
for cross-correlation of galaxy weak-lensing
maps and the CMB. An equivalent all-sky lensing treatment is
given by \cite{Ste97}. 

These quantities are given by the
second derivatives
\begin{eqnarray}
\nabla_i \nabla_j \len &\equiv&
	\kappa g_{ij} + (\gamma_1 + i \gamma_2)\Mp_{ij}  \nonumber\\
	&& \quad + (\gamma_1 - i \gamma_2)\Mn_{ij}\,,
\end{eqnarray}
 convergence 
where $g_{ij}$ is the metric on the sphere.

For the all-sky harmonics, it is useful to note that equation
(\ref{eqn:raiselower}) implies
\begin{eqnarray}
\nabla_i \nabla_j Y_l^m &=& 
	- {l(l+1) \over 2}Y_l^m g_{ij} 
	+ {1 \over 2}
	\sqrt{ (l+2)! \over (l-2)!} 
\\
&&\times	\Big[ \spin{2} Y_l^m \Mp
	+ \!\spin{-2} Y_l^m \Mn \Big]_{ij}
	\,,\nonumber
\end{eqnarray}
and hence 
\begin{eqnarray}
\kappa(\bn) &=& - \sum_{l m} {1 \over 2}l (l+1) \len_{l m} Y_l^m(\bn)\,, \nonumber\\
\gamma_1(\bn) \pm i \gamma_2(\bn) 
       &=& \sum_{l m} {1 \over 2} \sqrt{ (l+2)! \over (l-2)! } \len_{l m}\spin{\pm 2} Y_l^m(\bn) \,.
\end{eqnarray}
Consequently, the power spectra are related as
\begin{eqnarray}
C_l^{\kappa\kappa} &=& {l^2(l+1)^2 \over 4} C_l^{\phi\phi}\,, \nonumber\\
C_l^{\epsilon\epsilon} &=& {1 \over 4} {(l+2)! \over (l-2)!} C_l^{\phi\phi}  
\nonumber\,,\\
C_l^{X\kappa} &=& -{1 \over 2} l(l+1) C_l^{X\len} \,,\nonumber\\
C_l^{X\epsilon} &=& {1 \over 2} \sqrt{(l+2)! \over (l-2)!} C_l^{X\len}\,,
\end{eqnarray}
where the $\epsilon$ shear power spectra is defined in the same way
as that of the $E$ polarization and $X=\cmb,E,B$.  
The $\beta$ shear power is the analogue of the
$B$ polarization power and vanishes for weak lensing. 

In the flat-sky limit, these expressions become
\begin{eqnarray}
\kappa(\bn) &=& -{1 \over 2}\intln l^2 \len(\l) e^{i \l \cdot \bn}\\
\gamma_1(\bn) \pm i \gamma_2(\bn)  &=&
	     -{1 \over 2}
		\intln l^2 \len(\l) e^{\pm 2i(\varphi_l-\varphi)}
		e^{i \l \cdot \bn} \,,\nonumber
\end{eqnarray} 
so that 
\begin{eqnarray}
C_{(l)}^{\kappa\kappa}&=& C_{(l)}^{\epsilon\epsilon} = {1 \over 4}l^4 C_{(l)}^{\phi\phi} \,, \nonumber\\
C_{(l)}^{X\kappa} &=& - C_{(l)}^{X\epsilon} = -{1 \over 2} l^2 C_{(l)}^{X\len}
\end{eqnarray}
These relations also give the 
 bispectrum of the shear and convergence
in terms of the bispectrum of the lensing potential
\begin{eqnarray}
B_{l_1 l_2 l_3}^{\kappa\kappa\kappa} &=&
{1 \over 8}  [l_1(l_1+1)l_2(l_2+1)l_3(l_3+1)]  
B_{l_1 l_2 l_3}^{\phi\phi\phi} \,,\nonumber\\
B_{l_1 l_2 l_3}^{\epsilon\epsilon\epsilon} &=&
{1 \over 8} \sqrt{ {(l_1+2)!\over(l_1-2)!} {(l_2+2)!\over(l_2-2)!}
	{(l_3+2)!\over(l_3-2)!}}
B_{l_1 l_2 l_3}^{\phi\phi\phi} \,,
\end{eqnarray}
with a similar relation for the flat-sky bispectra.

\section{Wigner-3$j$ Evaluation}
\label{sec:wigner}

\subsection{Exact Expressions}

The expressions for the power spectrum of the lensed temperature
and polarization distributions involve specific sets of 
Wigner-3$j$ symbols that can be efficiently evaluated.
The expression for the temperature involves, a set which has
a closed algebraic form: 
\begin{eqnarray}
\wj &=&
(-1)^{L/2} \frac{({L \over 2})!}{({L \over 2}-l_1)!({L \over 2}-l_2)!({L \over 2
}-l_3)!}
\\
\label{eqn:w3j000}
&\times&
\left[\frac{(L-2l_1)!(L-2l_2)!(L-2l_3)!}{(L+1)!}\right]^{1/2}\,,
\nonumber
\end{eqnarray}
for even $L=l_1+l_2+l_3$ and zero for odd $L$.   

The required set for the
polarization does not have an exact closed form expression.  However it may
be equally efficiently evaluated for our purposes with the realization that
in the sums, we require 
\begin{equation}
\wjma {l_1}{l_2}{l_3}{m_1}{m_2}{m_3}  \equiv w_{l_1}
\end{equation}
for fixed $l_2,l_3,m_1,m_2,m_3$ and all allowed $l_1$. 
The recursion relations for the Wigner-$3j$ symbol,
\begin{eqnarray}
l_1 A_{l_1+1} w_{l_1 +1} + B_{l_1} w_{l_1} + (l_1+1) A_{l_1} w_{l_1-1}= 0 \,,
\end{eqnarray}
where
\begin{eqnarray}
A_{l_1} &=& \sqrt{l_1^2 - (l_2 - l_3)^2}
	    \sqrt{(l_2 + l_3 + 1)^2 - l_1^2} 
	    \sqrt{l_1^2 - m_1^2}\,, \nonumber\\
B_{l_1} &=& -(2 l_1 +1) [l_2(l_2+1)m_1 - l_3(l_3+1)m_1 
	\nonumber \\&& \quad
	- l_1(l_1+1)(m_3-m_2)] \,,
\end{eqnarray}
allow us to generate the whole set at once \cite{SchGor75}. 
For a stable recursion, one begins at the minimum and maximum $l_1$ values
\begin{eqnarray}
l_{1 {\rm min}} & = & \max(|l_2-l_3|),|m_1|)\,, \nonumber\\
l_{1 {\rm max}} & = & l_2 + l_3\,,
\end{eqnarray}
with $w_{l_{1 {\rm min}}} = w_{l_{1 {\rm max}}} = 1$ and carries the
recursion in both directions to the midpoint $l_{1 {\rm mid}}$ 
in the range (or any non-vanishing entry in the vicinity).
One then renormalizes either the left or right recursion to make
the $w_{l_{1 {\rm mid}}}$ agree.  The remaining overall normalization
is fixed by requiring
\begin{equation} 
\sum_{l_1} (2l_1+1) w_{l_1}^2 = 1 \,,
\end{equation}
and 
\begin{equation}
{\rm sgn}\left( w_{l_{1 \rm max}} \right)= (-1)^{l_2-l_3-m_1} \,.
\end{equation}
Putting these relations together, we obtain the full set of
symbols as required.
\begin{figure}[t]
\centerline{\epsfxsize=3.5truein\epsffile{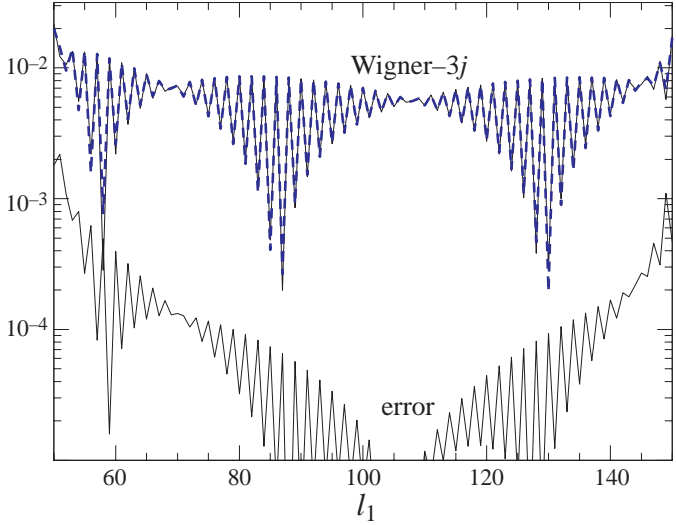}}
\caption{Wigner-3$j$ function and approximation.
An example of the Wigner-3$j$ symbol relevant to the polarization
calculation with $l_2=100$, $m_2=0$, $l_3=50$, $m_3=-2$ is shown
as calculated from the recursion relations (solid upper) and
analytic approximation (dashed).  The difference is shown below
(solid lower). 
}
\label{fig:w3j}
\end{figure}

\subsection{Approximations}

We can use the general relation between the all and flat sky bispectra
of equation~(\ref{eqn:correspondence}) compared with the explicit calculation
of the flat sky bispectrum in \S \ref{sec:bipol} 
to develop an high-$l$ approximation
for the specific symbol in the polarization calculations.  The comparison
implies that
\begin{equation}
\wjma{l_1}{l_2}{l_3}{2}{0}{-2} \approx
 	 \cos 2\varphi_{31}
	\wjma{l_1}{l_2}{l_3}{0}{0}{0}\,, 
\end{equation}
for $L=l_1+l_2+l_3=$ even.
By the law of cosines,
\begin{eqnarray}
\cos 2 \varphi_{31} = 
	{1 \over 2} 
				{(l_2^2 - l_1^2 - l_3^2)^2\over l_1^2 l_3^2} -1\,.
\end{eqnarray}	
Then 
\begin{eqnarray}
\wjma{l_1}{l_2}{l_3}{2}{0}{-2} &\approx&
	(-1)^{L/2} 
\left[ 
	{1 \over 2} {(l_2^2 - l_1^2 - l_3^2)^2\over l_1^2 l_3^2} -1\right]
\nonumber\\ &&\times
	{(L/2)! \over (L/2-l_1)!(L/2-l_2)!(L/2 -l_3)!}
\nonumber \\
&&\times
\left[\frac{(L-2l_1)!(L-2l_2)!(L-2l_3)!}{(L+1)!}\right]^{1/2} \,,
\nonumber
\end{eqnarray}
for $L=$ even.
For odd values of $L$, we use the relation 
\begin{equation}
\wjma{l_1}{l_2}{l_3}{2}{0}{-2} \approx
 	 \pm i \sin 2\varphi_{31}
	\wjma{l_1}{l_2}{l_3}{0}{0}{0}\,,  
\end{equation}
and fix the overall sign ambiguity by an explicit evaluation.
By the triangle relations,
\begin{eqnarray}
\sin 2\varphi_{31} &=& 
	\mp {1 \over 2} 
	   [ L(L -2l_1 )( L-2l_2 )	( L-2 l_3 ) ]^{1/2}
\nonumber\\
&&\quad \times
	   \left ( {l_2^2 - l_1^2 - l_3^2 \over  l_1^2  l_3^2} \right)\,.
\nonumber
\end{eqnarray}	
Putting this together with equation~(\ref{eqn:w3j000}) and fixing the
sign ambiguity, we obtain
\begin{eqnarray}
\wjma{l_1}{l_2}{l_3}{2}{0}{-2} &\approx& 
		(-1)^{L-1 \over 2} 
	   {1 \over 2}
		 \left ( {l_2^2 - l_1^2 - l_3^2 \over  l_1^2  l_3^2} \right)
\\ && \times
	{(L/2)! \over (L/2-l_1)!(L/2-l_2)!(L/2-l_3)!}
\nonumber\\ && \times
	   \Big[ L
		(L-2 l_1 )	
		(L-2 l_2 )	
		(L-2 l_3 )
\nonumber\\ && \times
{(L-2l_1)!(L-2l_2)!(L-2l_3)! \over (L+1)!}\Big]^{1/2}
\nonumber
\end{eqnarray}	
for $L={\rm odd}$.  The half integer factorials are defined by
the gamma function $x! = \Gamma(1+x)$.
By explicit calculation we find that these expressions are valid to
better than $3\%$ of the rms amplitude of the symbol when averaged
over neighboring $l$ for all 
$l_1 -|l_2-l_3| \agt 25$ 
and $l_2+l_3-l_1 \agt 25$, i.e. for triangles that are sufficiently
far from being flat. Near zero crossings,
the {\it fractional} error can be large but the absolute error
remains a small fraction of the rms.
A typical case is shown in Fig.~\ref{fig:w3j}.

These relations may be useful in cases where only a single symbol is
needed.   However for the lensing calculation where the whole
set is required, the recursion relations are as efficient as the
approximation and are exact.  

\section{Flat and All Sky Correspondence}  
\label{sec:correspondence}

\subsection{Harmonics}

We establish here the correspondence between the all and flat sky 
harmonic coefficients
of spin zero (scalar), spin one (vector) and spin two (tensor) quantities
on the sky.

Following \cite{WhiCarDraHol99}, let us begin by introducing the following
weighted sum over the multipole moments of the field $X=\Theta$, $E$, $B$,
or $\phi$ for a given $l$ and its inverse relation,
\begin{eqnarray}
X(\l) &=& \sqrt{4 \pi \over 2 l+1} \sum_m i^{-m} X_{l m} e^{i m \varphi_l}\,,\nonumber\\
X_{lm} &=& \sqrt{2 l+1 \over 4\pi} i^m \int {d \varphi_l \over 2\pi}
		e^{-i m\varphi_l} X(\l)\,.
\end{eqnarray}
The goal is then to show that this quantity  is the Fourier
coefficient of the flat-sky expansion.

Spin-0 quantities, such as the temperature flucutations and the
lensing potential, are decomposed as
\begin{equation}
X(\bn) = \sum_{lm} X_{lm} Y_l^m(\bn)\,.
\end{equation}
For small angles around the pole, the spherical harmonics
may be approximated as\footnote{Note
that our definition of $Y_l^m$ differs from the usual one by $(-1)^m$ 
to conform with the spin spherical harmonic convention \cite{SchGor75}.}
\begin{equation}
Y_l^{m} \approx J_m(l\theta) \sqrt{l \over 2\pi} e^{i m \varphi} \,,
\end{equation} 
and the expansion of the plane wave
\begin{eqnarray}
e^{i \l \cdot \bn} &=& \sum_m i^m J_m(l\theta) e^{im(\varphi -\varphi_l)} \nonumber\\
		   &\approx&
		\sqrt{2 \pi \over l}  \sum_m i^m Y_l^m e^{im \varphi_l} \,,
\label{eqn:besselplane}
\end{eqnarray}

Thus 
\begin{eqnarray}
X(\bn) &=& \sum_{l m} X_{lm} Y_l^m \nonumber\\
	    &\approx& 
		\sum_{l}
		{l \over 2\pi} 
		\int {d \varphi_l \over 2\pi}
		X(\l)
		\sum_{m} J_m(l\theta) 
		i^m e^{i m (\varphi-\varphi_l)}  \nonumber\\
	    &\approx&
		\int {d^2 l \over (2\pi)^2} X(\l) e^{i \l \cdot \bn}\,,
\label{eqn:xcorrespondence}
\end{eqnarray}	
which is the desired correspondence.

Spin-1 quantities like the deflection angles are decomposed as
\begin{equation}
{}_\pm X(\bn) = \sum_{l m} {}_\pm X_{lm} \spin{\pm 1} Y_l^m\,.
\end{equation} 
Here one notes that
\begin{equation} 
\spin{\pm 1} Y_l^m \approx \pm  {1 \over l} e^{\mp i\varphi}
		\left(\partial_x \pm i \partial_y\right) Y_l^m  \,,
\end{equation}
and thus
\begin{eqnarray}
{}_\pm X(\bn) &=& \sum_{l m} {}_\pm X_{lm} \spin{\pm 1} Y_l^m \nonumber\\
	    &\approx& 
		\pm \sum_{l}
		{l \over 2\pi} 
		\int {d \varphi_l \over 2\pi}
		{}_\pm X(\l) e^{\mp i\varphi}
		{1\over l}\left(\partial_x \pm i \partial_y\right) 
			e^{i \l \cdot \bn} \nonumber\\
	    &\approx&
		\pm i \int {d^2 l \over (2\pi)^2} {}_\pm X(\l) 
			e^{\pm i(\varphi_l -\varphi)}
			e^{i \l \cdot \bn}\,.
\label{eqn:spin1}
\end{eqnarray}

Finally, spin-2 quantities like the polarization are decomposed as
\begin{equation}
{}_\pm X(\bn) = \sum_{l m} {}_\pm X_{lm} \spin{\pm 2} Y_l^m\,.
\end{equation}
Here one notes that
\begin{equation} 
\spin{\pm 2} Y_l^m \approx {1 \over l^2} e^{\mp 2i\varphi}
		\left(\partial_x \pm i \partial_y\right)^2 Y_l^m  \,,
\end{equation}
and thus
\begin{eqnarray}
{}_\pm X(\bn) &=& \sum_{l m} {}_\pm X_{lm} \spin{\pm 2} Y_l^m \nonumber\\
	    &\approx& 
		\sum_{l}
		{l \over 2\pi} 
		\int {d \varphi_l \over 2\pi}
		{}_\pm X(\l) e^{\mp 2i\varphi}
		{1\over l^2}\left(\partial_x \pm i  \partial_y \right)^2 
			e^{i \l \cdot \bn} \nonumber\\
	    &\approx&
		-\int {d^2 l \over (2\pi)^2} {}_\pm X(\l) 
			e^{\pm 2i(\varphi_l -\varphi)}
			e^{i \l \cdot \bn}\,,
\end{eqnarray}
as desired.

\subsection{Power Spectra}

The correspondence between power spectra then follows from the 
relationship between the harmonics
\begin{eqnarray}
\left< X_{l m}^* X_{l' m'}' \right>
	&\approx&
		i^{m'-m} 
		{\sqrt{l l'}\over 2\pi}
		C^{XX'}_{(l)}  
	  \int d \varphi_l 
		e^{im\varphi_l}
	  \int d \varphi_{l'} 
		e^{-im' \varphi_{l'}}
\nonumber\\&&\quad\times
		\delta(\l-\l')  \,.
\end{eqnarray}
We then expand the delta function in plane waves functions
\begin{eqnarray}
\delta(\l - \l')& = & \int {d \bn \over (2\pi)^2} e^{i (\l - \l') \cdot \bn}   \\
\label{eqn:expanddelta2}
		&\approx& 
		{2 \pi \over \sqrt{ l l'}}  
		\int {d \bn \over (2\pi)^2} \sum_{m m'}
		i^{m-m'} 
		Y_{l'}^{m'*} 
		Y_l^m  
		e^{im \varphi_l - im'\varphi_{l'}} \,.\nonumber
\end{eqnarray}
Integrating over the azimuthal 
angles $\varphi_l,\varphi_{l'}$ collapses the sum to 
\begin{eqnarray}
\left< X_{l m}^* X_{l' m'}' \right> &\equiv&
\delta_{l,l'} \delta_{m,m'} C_l^{XX'}\nonumber\\
		&\approx& 
		C^{XX'}_{(l)}
			\int {d \bn} Y_{l}^{-m*} \, Y_{l'}^{-m'} 
			\nonumber\\
                &=& \delta_{l,l'} \delta_{m,m'} C^{XX'}_{(l)}
\end{eqnarray}
which proves the desired relation in eqn.~(\ref{eqn:correspondence}),
\begin{eqnarray}
C_l^{XX'} &\approx& C_{(l)}^{XX'} \,.
\end{eqnarray}

\subsection{Bispectra}

The correspondence between bispectra is established in exactly the
same way as with the power spectra.
The only difference is that the expansion of $\deld(\l'-\l)$ 
in eqn.~(\ref{eqn:expanddelta2}) is replaced
with that of $\deld(\l_1+\l_2+\l_3)$ 
leading to
\begin{eqnarray}
\left< X_{l m} X_{l' m'}' X_{l'' m''}'' \right> &\equiv&
\wjma{l}{l'}{l''}{m}{m'}{m''} B_{l l' l''}^{XX'X''}  
\\ &\approx &
	B^{XX'X''}_{(l,l',l'')}
			\int {d \bn} Y_{l}^{-m} \, Y_{l'}^{-m'}
				\, Y_{l''}^{-m''} \nonumber\\
\nonumber\\ &=& 
	B^{XX'X''}_{(l,l',l'')}
		\wjma{l}{l'}{l''}{0}{0}{0} 
		\wjma{l}{l'}{l''}{m}{m'}{m''}
\nonumber\\&& \quad\times
		\sqrt{(2l+1)(2l'+1)(2l''+1) \over 4\pi}  \,.
\nonumber
\end{eqnarray}
This establishes the relation
\begin{eqnarray}
B_{l l' l''}^{X X' X''} &\approx&  
			\wjma{l}{l'}{l''}{0}{0}{0}
			\sqrt{ (2l+1)(2l'+1)(2l''+1) \over 4\pi} \nonumber\\
	   &&\quad B^{X X' X''}_{(l,l',l'')}\,.
\label{eqn:bicorrespondence}
\end{eqnarray}
Note that we have implicitly assumed that 
the bispectrum only depends on the the magnitudes ($l,l',l''$) so that it may
be removed from the azimuthal integrals.  
This is not true for terms not involving the magnetic parity.
In this case, the sign of the flat-sky bispectrum depends on orientation
but we find empirically that a similar relationship holds up to a sign
ambiguity as discussed in \S \ref{sec:cmb}.

\subsection{Signal-to-Noise}

Here we establish the correspondence between the all and flat sky
signal-to-noise statistics for the case of diagonal
contributions to the covariance matrix 
[${\rm Cov}$ = {\rm diag}(${\rm Var}$)], 
\begin{eqnarray}
\left( { S \over N} \right)^2  &=& \sum_{l m} {(C_l^{XX})^2 \over {\rm Var}} 
=\sum_l (2l+1) 
			{(C_l^{XX})^2 \over {\rm Var}} \,.
\end{eqnarray}
For the flat sky case, one defines a weighted sum of Fourier harmonics
\begin{eqnarray}
P = \int d^2 l\, W(l) X(\l)X(-\l) \,,
\end{eqnarray}
with optimal weights given by $W(l)=C^{XX}_{(l)}/{\rm Var}$ from which one calculates
the signal-to-noise $\left< P \right>^2/\left< P^2 \right>$ as
\begin{eqnarray}
\left( { S \over N} \right)^2  &=& 
		{f_{\rm sky}\over \pi}\int d^2 l 
		{[C_{(l)}^{XX}]^2 \over {\rm Var}} \nonumber\\
			       &\approx& 
		 2 f_{\rm sky} \int l dl  {(C_l^{XX})^2 \over {\rm Var}},
\end{eqnarray}
where we have used the fact that $\deld({\bf 0})\approx V/(2\pi^2) = f_{\rm sky}/\pi$.
These expressions agree in the high $l$-limit and imply the familiar result that
$f_{\rm sky}$ should multiply the signal-to-noise of angular power spectrum
measurements given incomplete sky coverage.

The bispectrum signal-to-noise similarly is
\begin{eqnarray}
\left( { S \over N} \right)^2  &=& \sum_{l_1 l_2 l_3} 
				{(B_{l_1 l_2 l_3}^{XXX})^2 \over {\rm Var}} 
				\sum_{m_1 m_2 m_3}
				\wjma{l_1}{l_2}{l_3}{m_1}{m_2}{m_3}^2 
				\nonumber\\
			      &=& \sum_{l_1 l_2 l_3} 
				{(B_{l_1 l_2 l_3}^{XXX})^2 \over {\rm Var}} \,,
\end{eqnarray}
for the all sky bispectrum and
\begin{eqnarray}
\left( { S \over N} \right)^2  &=& 
		{f_{\rm sky}\over \pi}{1 \over (2\pi)^2} \int d^2 l_1
		\int d^2 l_2
		{[B_{(l_1,l_2,l_3)}^{XXX}]^2 \over {\rm Var}} \,,
\end{eqnarray}
for the flat sky bispectrum \cite{Zal99}.
The extra factor of $(2\pi)^2$ compared with the power 
spectrum is from the extra delta function in the noise term.
One can show that these expressions agree in the high-$l$ limit by restoring
the integration over $l_3$, expanding the delta function into spherical
harmonics as in eqn.~(\ref{eqn:expanddelta2}), and integrating over
azimuthal angles,
\begin{eqnarray}
&&
\int d^2 l_1 \int d^2 l_2 \int d^2 l_3 \deld (\l_1+\l_2+\l_3) \nonumber\\
&&\quad  \approx \int l_1 dl_1 \int l_2 dl_2 \int l_3 d l_3
	{\sqrt{ (2\pi)^5 \over l_1 l_2 l_3}} \int d\bn Y_{l_1}^0 Y_{l_2}^0 Y_{l_3}^0 
	\nonumber\\
&&\quad \approx 8\pi^2 \int l_1 dl_1 \int l_2 dl_2 \int l_3 d l_3 
	\wjma{l_1}{l_2}{l_3}{0}{0}{0}^2 \,.
\end{eqnarray} 
With the general correspondence of bispectra from eqn.~(\ref{eqn:bicorrespondence}),
this becomes
\begin{equation}
\left( { S \over N} \right)^2  \approx 
		f_{\rm sky}
		\int dl_1 \int dl_2 \int dl_3 
				{(B_{l_1 l_2 l_3}^{XXX})^2 \over {\rm Var}} \,,
\end{equation}
which proves the equivalence of the signal-to-noise for high-$l$ and $f_{\rm sky}=1$.


\begin{thebibliography}{99}

\bibitem{PreRef} 
		 A. Blanchard \& J. Schneider, Astron. \& Astrophys., 184, 1 (1987);
		 A. Kashlinsky, Astrophys. J. 331, L1 (1988);
		 E.V. Linder, Astron. \& Astrophys. 206, 199 (1988);
		 S. Cole \& G. Efstathiou, Mon. Not. Roy. Astron. Soc. 239, 195 (1989);
		 M. Sasaki, Mon. Not. Roy. Astron. Soc., 240, 415 (1989);
		 K. Watanabe \& K. Tomita, Astrophys. J. 370, 481 (1991)
		 M. Fukugita, T. Futamase, M. Kasai, \& E.L. Turner, Astrophys. J. 393, 3 (1992);
		 L. Cayon, E. Martinez-Gonzalez, \& Sanz, Astrophys. J., 413, 10 (1993);

\bibitem{Sel96}
	U. Seljak, Astrophys. J, 463 1 (1996)

\bibitem{ZalSel98}
        M. Zaldarriaga \& U. Seljak, Phys. Rev. D. 58, 023003 (1998)


\bibitem{GolSpe99}
        D.M. Goldberg \& D.N. Spergel, Phys. Rev. D, 59, 103002 (1999)

\bibitem{Zal99}
	M. Zaldarriaga, preprint, astro-ph/9910498

\bibitem{SelZal96}
        U. Seljak \& M. Zaldarriaga,  Astrophys. J., 469, 437 (1996)

\bibitem{EisHu99}
        D.J. Eisenstein \& W. Hu, Astrophys. J., 511, 5 (1999)

\bibitem{BunWhi97}
        E.F. Bunn \& M. White, Astrophys. J., 480, 6 (1997)

\bibitem{Pee80}
        Peebles, P.J.E. 1980, The Large-Scale Structure of the Universe,
        (Princeton: Princeton Univ. Press)

\bibitem{SelZal99}
        U. Seljak \& M.  Zaldarriaga,
        preprint, astro-ph/9811123

\bibitem{Goletal67}
        J.N. Goldberg, et al. J. Math. Phys., 7, 863 (1967)

\bibitem{KamKosSte97}
	M. Kamionkowski, A. Kosowsky, \& A. Stebbins,
        Phys. Rev. D, 55, 7368 (1997)

\bibitem{ZalSel97}
	M. Zaldarriaga \& U. Seljak, Phys. Rev. D., 55, 1830 (1997)

\bibitem{Kai98}
        N. Kaiser, Astrophys. J., 498, 26 (1998)

\bibitem{BarSch99}
	M. Bartlemann \& P. Schneider, preprint, astro-ph/9912508

\bibitem{PeaDod96}
        J.A. Peacock \& S.J. Dodds,  Mon. Not. Roy. Astron. Soc., 280, L19 (1996)




\bibitem{HuWhi97}
	W. Hu \& M. White, Phys. Rev. D, 56 596 (1997)

\bibitem{Luo94}
	X. Luo, Astrophys. J. 427, 71 (1994)


\bibitem{CooHu99} A. Cooray \& W. Hu, preprint, astro-ph/9910397


\bibitem{Tegetal99}
        M. Tegmark, D.J. Eisenstein, W. Hu, \& A.  de Oliveira-Costa, 
        Astrophys. J, in press (2000)

		
\bibitem{Ste97}
	A. Stebbins, preprint, astro-ph/9609149





\bibitem{SchGor75}
        K. Schulten \& R.G. Gordon, J. Math. Phys., 16, 1971 (1975)






\bibitem{WhiCarDraHol99}
        M. White, J.E. Carlstrom, M. Dragovan,  \& W.L. Holzapfel, W.L. 
        Astrophys. J., 514, 12 (1999)



\end{thebibliography}
\end{document}